\pdfoutput=1  %
\documentclass[11pt,a4paper]{article}

\usepackage[utf8]{inputenc}
\usepackage[T1]{fontenc}
\usepackage{lmodern}
\usepackage[margin=2.5cm]{geometry}
\usepackage{amsmath,amssymb,bm,mathtools}
\usepackage{graphicx}
\usepackage{booktabs}
\usepackage{siunitx}
\usepackage{xcolor}
\usepackage[backend=biber,style=numeric-comp,sorting=none,natbib=true]{biblatex}
\usepackage{caption}
\usepackage{subcaption}
\usepackage{tikz}
\usetikzlibrary{arrows.meta}
\usepackage{pgfplots}
\pgfplotsset{compat=1.18}
\usepgfplotslibrary{groupplots}
\usepackage[title]{appendix}  %
\usepackage[colorlinks=true,linkcolor=blue!50!black,citecolor=blue!50!black,urlcolor=blue!50!black]{hyperref}
\usepackage[capitalise]{cleveref}
\usepackage[section]{placeins}  %

\newif\ifshowtodos \showtodostrue

\showtodosfalse  %
\newcommand{\uu}{\bm{u}}
\newcommand{\ww}{\bm{w}}
\newcommand{\nn}{\bm{n}}
\newcommand{\xx}{\bm{x}}
\newcommand{\Q}{\mathcal{Q}}
\newcommand{\Pst}{\mathcal{P}}
\newcommand{\dd}{\,\mathrm{d}}
\newcommand{\Rb}{R_{\mathrm{b}}}

\DeclareSIUnit\mmHg{mmHg}
\DeclareSIUnit\mL{mL}
\DeclareSIUnit\dyn{dyn}

\definecolor{primary}   {RGB}{  0  84 159}
\definecolor{primary-75}{RGB}{ 64 127 183}
\definecolor{primary-50}{RGB}{142 186 229}
\definecolor{primary-25}{RGB}{199 221 242}
\definecolor{primary-10}{RGB}{232 241 250}

\definecolor{black}   {RGB}{  0   0   0}
\definecolor{black-75}{RGB}{100 101 103}
\definecolor{black-50}{RGB}{156 158 159}
\definecolor{black-25}{RGB}{207 209 210}
\definecolor{black-10}{RGB}{236 237 237}

\definecolor{magenta}   {RGB}{227   0 102}
\definecolor{magenta-75}{RGB}{233  96 136}
\definecolor{magenta-50}{RGB}{241 158 177}
\definecolor{magenta-25}{RGB}{249 210 218}
\definecolor{magenta-10}{RGB}{253 238 240}

\definecolor{yellow}   {RGB}{255 237   0}
\definecolor{yellow-75}{RGB}{255 240  85}
\definecolor{yellow-50}{RGB}{255 245 155}
\definecolor{yellow-25}{RGB}{255 250 209}
\definecolor{yellow-10}{RGB}{255 253 238}

\definecolor{petrol}   {RGB}{  0  97 101}
\definecolor{petrol-75}{RGB}{ 45 127 131}
\definecolor{petrol-50}{RGB}{125 164 167}
\definecolor{petrol-25}{RGB}{191 208 209}
\definecolor{petrol-10}{RGB}{230 236 236}

\definecolor{turkis}   {RGB}{  0 152 161}
\definecolor{turkis-75}{RGB}{  0 177 183}
\definecolor{turkis-50}{RGB}{137 204 207}
\definecolor{turkis-25}{RGB}{202 231 231}
\definecolor{turkis-10}{RGB}{235 246 246}

\definecolor{grun}   {RGB}{ 87 171  39}
\definecolor{grun-75}{RGB}{141 192  96}
\definecolor{grun-50}{RGB}{184 214 152}
\definecolor{grun-25}{RGB}{221 235 206}
\definecolor{grun-10}{RGB}{242 247 236}

\definecolor{maigrun}   {RGB}{189 205   0}
\definecolor{maigrun-75}{RGB}{208 217  92}
\definecolor{maigrun-50}{RGB}{224 230 154}
\definecolor{maigrun-25}{RGB}{240 243 208}
\definecolor{maigrun-10}{RGB}{249 250 237}

\definecolor{orange}   {RGB}{246 168   0}
\definecolor{orange-75}{RGB}{250 190  80}
\definecolor{orange-50}{RGB}{253 212 143}
\definecolor{orange-25}{RGB}{254 234 201}
\definecolor{orange-10}{RGB}{255 247 234}

\definecolor{rot}   {RGB}{204   7  30}
\definecolor{rot-75}{RGB}{216  92  65}
\definecolor{rot-50}{RGB}{230 150 121}
\definecolor{rot-25}{RGB}{243 205 187}
\definecolor{rot-10}{RGB}{250 235 227}

\definecolor{bordeaux}   {RGB}{161  16  53}
\definecolor{bordeaux-75}{RGB}{182  82  86}
\definecolor{bordeaux-50}{RGB}{205 139 135}
\definecolor{bordeaux-25}{RGB}{229 197 192}
\definecolor{bordeaux-10}{RGB}{245 232 229}

\definecolor{violett}   {RGB}{ 97  33  88}
\definecolor{violett-75}{RGB}{131  78 117}
\definecolor{violett-50}{RGB}{168 133 158}
\definecolor{violett-25}{RGB}{210 192 205}
\definecolor{violett-10}{RGB}{237 229 234}

\definecolor{lila}   {RGB}{122 111 172}
\definecolor{lila-75}{RGB}{155 145 193}
\definecolor{lila-50}{RGB}{188 181 215}
\definecolor{lila-25}{RGB}{222 218 235}
\definecolor{lila-10}{RGB}{242 240 247}

\pgfplotsset{
  clean/.style={
    axis lines=left, axis line style={black-50}, tick style={black-50},
    tick label style={font=\scriptsize, color=black-75},
    label style={font=\scriptsize, color=black-75},
    every axis plot/.append style={line join=round},
  },
}
\colorlet{model}{primary}
\colorlet{measured}{black-50}
\colorlet{systole}{orange-10}
\newcommand\panelcaption[4]{%
  \node[anchor=north, text width=#2, inner sep=0pt, yshift=-1.5mm] at (#1) {\subcaption{#3}\label{#4}};}

\title{Continuous Space-Time Finite Element Simulations for Biomedical Flows with Large Boundary Motion and Time-Periodicity}
\author{ 
  Nico Dirkes\textsuperscript{*}, Marek Behr
  \\[1em]
  {\small Chair for Computational Analysis of Technical Systems}
  \\
  {\small RWTH Aachen University, Germany}
  \\[1em]
  {\small\textsuperscript{*}Corresponding author}
}
\date{\today}

\begin{document}
\maketitle

\begin{abstract}
We present an axisymmetric continuous simplex space-time finite element method for incompressible fluid flow. Compared to conventional time-stepping methods, the space-time approach has several advantages. First, the motion of the domain is incorporated into the space-time geometry, allowing for large boundary motion without remeshing or projection of the solution. Second, the method allows for time-periodic solutions to be computed directly, without the need to simulate multiple cycles. Third, the space-time mesh can be partitioned in time as well as in space, which enables parallelism in time. The method is demonstrated on an intra-aortic balloon pump (IABP) in a rigid tube. The balloon inflation is modelled based on measured data, and includes a travelling wave along the balloon. The method is able to resolve complex flow features such as jets and recirculation zones. A refinement study shows convergence under temporal, spatial and combined space-time refinement. In weak scaling tests up to 960 ranks, the cost per solver iteration remains approximately constant, and partitioning the mesh into time slabs improves the convergence of the linear solver compared to partitions that cut through both space and time.
\end{abstract}

\noindent\textbf{Keywords:} computational fluid dynamics; haemodynamics; intra-aortic balloon pump; continuous simplex space-time finite elements; high-performance computing

\section{Introduction}
\label{sec:intro}

Computational fluid dynamics (CFD) has become an important tool for the design and analysis of biomedical devices. Biomedical flows often involve complex geometries and moving boundaries, either through moving mechanical components or the motion of the vessel walls. Examples of this are ventricular assist devices, heart valves, and aneurysm stents. The motion of the boundaries can be large and complex, and it is often periodic in nature. Accurately capturing the flow features in these devices is crucial for their performance and safety.

Most commercial CFD codes are based on finite volume methods, or finite element methods with time-stepping. In these methods, the time discretisation is performed using finite differences, and the solution is advanced in time step by step. 
The most common approach to handle moving boundaries in this setting is the Arbitrary Lagrangian-Eulerian (ALE) method~\cite{Hughes1981,Donea1982}. Thereby, the boundaries are moved according to the prescribed motion. The computational mesh inside the domain is updated accordingly. This update can be performed by two main strategies: elastic mesh update or remeshing. In the elastic mesh update method, the mesh is deformed according to the boundary motion using a spring analogy. This method works well for small to moderate deformations, but can lead to poor mesh quality for large deformations. In the remeshing approach, a new mesh is generated whenever the mesh quality falls below a certain threshold. The solution from the old mesh is then projected onto the new mesh. This approach can handle larger deformations, but it introduces additional computational cost and potential errors due to the projection. The two strategies can also be combined, where the mesh is elastically updated until a certain quality threshold is reached, and then remeshing is performed. Alternative approaches include more sophisticated mesh update methods %
or immersed boundary methods.

However, finite element methods offer additional flexibility in the choice of time discretisation. In particular, \emph{space-time finite elements} treat time as an additional dimension, and the solution is discretised in both space and time using finite elements~\cite{jametNumericalSolutionEulerian1975,lynchFiniteElementSimulation1980}.
This allows for a more natural treatment of moving boundaries, as the motion can be incorporated into the space-time geometry~\cite{Tezduyar1992a,Tezduyar1992b,hansboVelocitypressureStreamlineDiffusion1990}. Most early works have used the concept of space-time slabs, which are temporal partitions of the space-time domain. Each subdomain is then solved sequentially, imposing weak continuity in time. This limits the size of the linear systems to be solved. 
In contrast, continuous space-time methods solve for the whole time period at once, featuring a fully unstructured mesh~\cite{Hughes1988,maubachIterativeMethodsNonlinear1994,Behr2008}. This yields higher flexibility in mesh generation, particularly with respect to mesh refinement in space and time~\cite{kornmaierAdaptiveRefinementSimplex2026}. Additionally, it leads to larger linear systems, enabling higher degrees of parallelisation. For further comparison between the different types of space-time methods, see~\cite{hostersAddressingMajorBottleneck2025}.
Time-continuous space-time methods on fully unstructured meshes have been applied to parabolic problems~\cite{Steinbach2015}, advection-diffusion problems~\cite{vonDanwitz2023}, and rotating electric machines~\cite{Gangl2024}. More recently, this approach has been applied to incompressible Navier-Stokes~\cite{Karyofylli2022,kharaSolvingFluidFlow2026,wegeMarkingStrategiesAdaptive2026a}. However, the field of biomedical flows is still dominated by time-discontinuous space-time or time-stepping methods~\cite{Pauli2017,africaLifexcfdOpensourceComputational2024,quarteroniCardiovascularSystemMathematical2017}.

In this study, we present a new application of continuous space-time finite elements. Specifically, we focus on the flow around an intra-aortic balloon pump (IABP), a device used to support the heart in patients with cardiac failure~\cite{Moulopoulos1962,Kantrowitz1968}. The IABP consists of a balloon that is inserted into the aorta through a catheter. The balloon is synchronised with the cardiac cycle. When the aortic valve is closed during diastole, the balloon is inflated, which displaces blood towards the coronary arteries. Right before the aortic valve opens, the balloon is deflated, which creates a suction effect that draws blood from the left ventricle into the aorta. This process is known as diastolic counterpulsation. Most commonly, the IABP has been modelled using lumped-parameter models, which treat the balloon as a volume or pressure source~\cite{Jaron1983,Abdolrazaghi2010,Schampaert2013,DeLazzari2020} and resolve its interaction with the cardiovascular system using ordinary differential equations (ODEs). One-dimensional models have been used to study the propagation of pressure waves in the aorta~\cite{Kuklinski1984,Niederer1988,KimOhley1992}. However, these methods do not resolve the flow field around the balloon. More recently, three-dimensional CFD simulations have been performed~\cite{Caruso2016,Gu2019,Ma2025,Moulinos2023}. The balloon motion in these works is prescribed as a time-dependent global radius. However, recent experiments have shown that balloon motion is often more complex and non-uniform~\cite{Bruti2015}. A segmented balloon model~\cite{Bruti2016} has incorporated this effect, but only for three segments of piece-wise constant radius. 

We use this device to demonstrate the capabilities of continuous space-time finite elements for biomedical flows with large boundary motion. We employ a fully unstructured space-time mesh, which allows us to accurately capture the complex flow features around the balloon. We implement a more detailed balloon motion law, which is based on experimental measurements of the balloon radius at multiple locations along its length. We resolve the inflation front travelling along the balloon and study its effect on the flow field. Further, we employ periodic boundary conditions in time to directly compute the periodic state of the flow. Finally, we observe parallel scalability of the method under mesh refinement, which is crucial for the application of this method to more complex geometries and flow conditions.

The remainder of this paper is organised as follows. In \cref{sec:method}, we present the governing equations and the numerical method used in this study. In \cref{sec:numerical_setup}, we describe the configuration of our application case, including the geometry, boundary conditions, and mesh generation. In \cref{sec:results}, we present the results of our simulations and show timings for the parallel scalability of the method. Finally, in \cref{sec:conclusions}, we discuss the implications of our findings and provide an outlook for future work.

\section{Governing equations and space-time discretisation}
\label{sec:method}

The governing equations are the incompressible Navier--Stokes equations for a Newtonian fluid with density $\rho$ and dynamic viscosity $\mu$. In a moving domain $\Omega(t)\subset\mathbb{R}^n$, $n\in\{2,3\}$, $t\in(0,T)$, with Dirichlet and Neumann boundary conditions on $\Gamma_D(t)$ and $\Gamma_N(t)$, respectively, the strong form reads:
\begin{equation}
\begin{alignedat}{2}
  \bm{\mathcal R}_{\mathrm{MOM}}(\uu,p) \coloneqq \rho\left(\partial_t\uu + \uu\cdot\nabla\uu\right) - \nabla\cdot\bm{\sigma}(\uu,p) &= \bm{0}
  \qquad && \text{in } \Omega(t)\, , \\
  \bm{\sigma}(\uu,p) = -p\bm{I} + 2\mu\,\bm{\varepsilon}(\uu),
  &\qquad \bm{\varepsilon}(\uu) = \tfrac12\left(\nabla\uu+\nabla\uu^\top\right)  \, , \qquad  && \text{in } \Omega(t)\, ,\\
  \mathcal R_{\mathrm{CONT}}(\uu) \coloneqq \nabla\cdot\uu &= 0
  \qquad && \text{in } \Omega(t)\, , \\
  \uu &= \bm g && \text{on } \Gamma_D(t) \, , \\
  \bm{\sigma}\nn &= \bm h && \text{on } \Gamma_N(t) \, . 
\end{alignedat} \label{eq:ns_strong}
\end{equation}

In this work, we utilise cylindrical coordinates $(r,\theta,z)$ and assume axisymmetry, i.e., $\partial_\theta\equiv0$ and no swirl, $u_\theta=0$. The velocity vector is then $\uu=(u_r,u_z)$ and the three-dimensional spatial domain is restricted to two dimensions, $\Omega(t)\subset\{(r,z):r\ge0\}$. For this case, the governing equations \eqref{eq:ns_strong} can be expressed in terms of the axisymmetric operators:
\begin{equation}
\begin{aligned}
  \nabla\cdot\uu &= \frac1r\,\partial_r(r u_r) + \partial_z u_z \, , \\
  \uu\cdot\nabla\uu &=
  \begin{pmatrix}
    u_r\,\partial_r u_r + u_z\,\partial_z u_r\\
    u_r\,\partial_r u_z + u_z\,\partial_z u_z
  \end{pmatrix} \, ,\\
  \bm{\varepsilon}(\uu) &=
  \begin{pmatrix}
    \partial_r u_r & 0 & \tfrac12\left(\partial_z u_r + \partial_r u_z\right)\\
    0 & u_r/r & 0\\
    \tfrac12\left(\partial_z u_r + \partial_r u_z\right) & 0 & \partial_z u_z
  \end{pmatrix}_{(r,\theta,z)} \, , \\
  \nabla\cdot\bm{\sigma} &=
  \begin{pmatrix}
    \frac1r\,\partial_r(r\sigma_{rr}) + \partial_z\sigma_{rz} - \frac{\sigma_{\theta\theta}}{r}\\[2pt]
    \frac1r\,\partial_r(r\sigma_{rz}) + \partial_z\sigma_{zz}
  \end{pmatrix} \, .
\end{aligned} \label{eq:axi_operators}
\end{equation}
In order to ensure that the velocity field is well-defined at $r=0$, we require an additional boundary condition on the axis of symmetry $\Gamma_a=\{(r,z):r=0\}$:
\begin{equation}
  u_r = 0 \qquad \text{on } \Gamma_a \, . \label{eq:bc_axis}
\end{equation}

\subsection{Space-time finite element formulation}
\label{sec:weak}
\begin{figure}[htbp]
  \centering
  \input{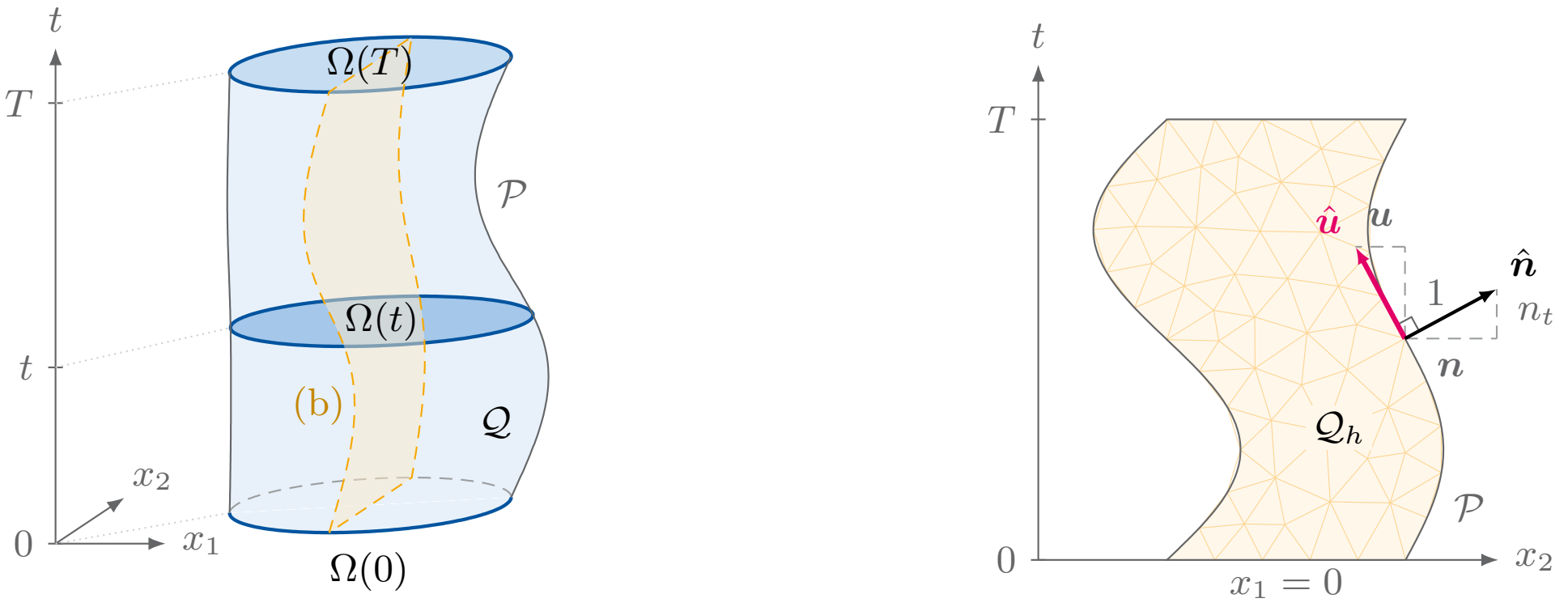}
  \caption{Illustration of the space-time domain. 
  }
  \label{fig:st_sketch}
\end{figure}

For the space-time finite element formulation, we define the space-time domain $\Q$ as
\begin{equation}
  \Q = \{ (\xx,t): \xx\in\Omega(t), t\in(0,T)\} \subset \mathbb{R}^{n+1} \, ,
\end{equation}
and its boundary $\Pst = \partial \Q$ with Dirichlet portions $\Pst_D$, Neumann portions $\Pst_N$ and axisymmetric boundary portions $\Pst_a$. The ansatz and test spaces are
\begin{equation}
\begin{gathered}
  {\mathcal S}_{\uu} = \left\{\uu\in[H^1(\Q)]^n : \uu=\bm g \text{ on } \Pst_D\right\} \, , \qquad
  {\mathcal V}_{\uu} = \left\{\ww\in[H^1(\Q)]^n : \ww=\bm 0 \text{ on } \Pst_D\right\} \, ,\\
  \mathcal S_p = \mathcal V_p = L^2(\Q) \, .
\end{gathered}
  \label{eq:spaces}
\end{equation}
The superscript $h$ denotes their discrete counterparts: continuous, piecewise linear functions on a partition $\Q_h=\{\Q^e\}_{e=1}^{n_{el}}$ of $\Q$, equal order for velocity and pressure (\cref{fig:st_sketch}).
 We apply Galerkin least-squares (GLS) stabilisation to enable the use of equal-order interpolation. The weak form of the axisymmetric incompressible Navier--Stokes equations \eqref{eq:ns_strong} reads:

Find $(\uu_h,p_h)\in{\mathcal S}_{\uu}^h\times\mathcal S_p^h$ such that for all $(\ww_h,q_h)\in{\mathcal V}_{\uu}^h\times\mathcal V_p^h$:
\begin{multline}
  \int_\Q \Big[\ww_h\cdot\rho\left(\partial_t\uu_h + \uu_h\cdot\nabla\uu_h\right)
    + \bm\varepsilon(\ww_h):2\mu\,\bm\varepsilon(\uu_h)
    - p_h\,\nabla\cdot\ww_h + q_h\,\nabla\cdot\uu_h\Big]\, r\dd\Q \\
  + \sum_{e=1}^{n_{el}} \int_{\Q^e}
    \frac{\tau_{\mathrm{MOM}}}{\rho}
    \Big[\rho\left(\partial_t\ww_h + \uu_h\cdot\nabla\ww_h\right) + \nabla q_h\Big]\cdot\bm{\mathcal R}_{\mathrm{MOM}}(\uu_h,p_h)\; r\dd\Q \\
  + \sum_{e=1}^{n_{el}} \int_{\Q^e}
    \rho\,\tau_{\mathrm{CONT}}\,\nabla\cdot\ww_h\;\mathcal R_{\mathrm{CONT}}(\uu_h)\; r\dd\Q
  = \int_{\Pst_N} \ww_h\cdot\bm h\; r\dd\Pst ,
  \label{eq:weak}
\end{multline}
with the residuals from \cref{eq:ns_strong} and all differential operators from \cref{eq:axi_operators}.
The first line is the Galerkin part; the second and third lines are the GLS stabilisation of the momentum and continuity equations with the space-time element metric \citep{Pauli2017}. The stabilisation parameters are computed from the covariant metric tensor $\bm G$ of the space-time element mapping, see \citet{pauliStabilizedFiniteElement2016}.
For linear elements, the viscous part of $\bm{\mathcal R}_{\mathrm{MOM}}(\uu_h,p_h)$ does not vanish. We therefore evaluate $\nabla\cdot\bm\sigma$ in the residual with velocity gradients recovered at the nodes by an $L^2$ projection with lumped mass matrix.

The problem is formulated over one period without an initial condition. Time periodicity, $\uu(\xx,0)=\uu(\xx,T)$ and $p(\xx,0)=p(\xx,T)$, is imposed strongly, see \cref{sec:numerical_setup}. In contrast to time-slab formulations, there is therefore neither an initial condition nor a temporal jump term in \eqref{eq:weak}.
The flow through the open boundaries $\Gamma_N$ changes direction over the cycle. Where fluid enters through a traction boundary, the convective term introduces kinetic energy through the boundary, which can cause divergence. Following \citet{esmailymoghadamComparisonOutletBoundary2011}, we therefore prescribe the traction
\begin{equation}
  \bm h = -p_k(t)\,\nn + \beta\rho\,(\uu\cdot\nn)_-\,\uu,
  \qquad (\uu\cdot\nn)_- = \min(\uu\cdot\nn,0) \, ,
  \label{eq:backflow}
\end{equation}
with the prescribed pressure $p_k(t)$ on each open boundary $k$. The additional term vanishes where fluid flows out. The term is applied on both open boundaries, as either can see inflow and outflow over the cycle. The parameter $\beta$ is set to $1/2$.

\clearpage
\subsection{Kinematic condition}
\label{sec:kinematic}
At moving walls, the fluid velocity must satisfy the kinematic condition, which states that the normal component of the fluid velocity at the wall must equal the normal component of the wall velocity. 
However, the space-time mesh is generated in advance and does not move. The boundary motion is contained within the geometry of $\Pst$. We therefore derive the kinematic condition from the space-time domain. Let $\hat\nn=(\nn,n_t)$ be the outward space-time normal of $\Pst$, split into a spatial part $\nn$ and a time component $n_t$. A fluid particle advances in space-time along $\hat\uu=(\uu,1)$. The kinematic condition in space-time then reads
\begin{equation}
  \hat\uu\cdot\hat\nn = 0
  \qquad\Longleftrightarrow\qquad
  \uu\cdot\nn = -n_t \, .
  \label{eq:vn}
\end{equation}
In practice, this means that the normal velocity $\uu\cdot\nn$ is determined from the time component $n_t$ of the space-time normal $\hat\nn$. On fixed walls, $n_t=0$ and \eqref{eq:vn} reduces to the no-penetration condition.
This condition is illustrated in \cref{fig:st_sketch_cut}. We note that only the normal velocity can be determined this way; the tangential velocity is prescribed separately. We evaluate $\hat\nn$ on the boundary of the discrete space-time mesh, so the wall velocity follows from the mesh alone, without the need for an analytic description of the motion.

\section{Computational model}
\label{sec:numerical_setup}
We demonstrate the continuous simplex space-time approach from \cref{sec:method} on an intra-aortic balloon pump (IABP) test case. In this section, we describe the geometry and boundary conditions for this test case, as well as the space-time mesh generation.

\subsection{Geometry and motion}
\label{sec:geometry}
\begin{figure}
  \centering
  \begin{subfigure}{\linewidth}
    \centering
    \input{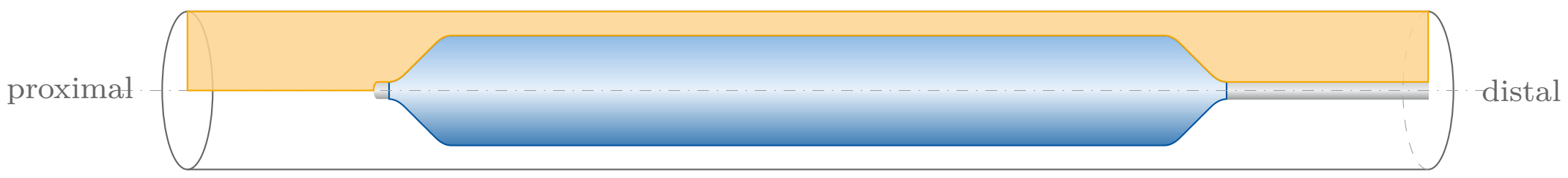}
    \caption{Tube with the inflated balloon and the meridional half-plane (orange).}
    \label{fig:geometry:3d}
  \end{subfigure}
  \\[2ex]
  \begin{subfigure}{\linewidth}
    \centering
    \input{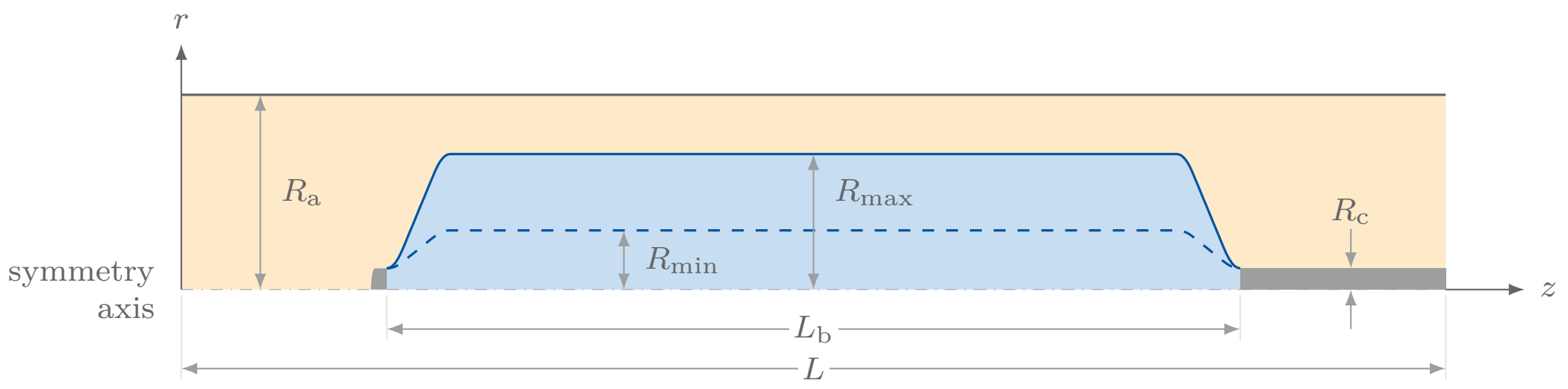}
    \caption{Axisymmetric domain with the inflated (solid) and deflated (dashed) balloon.}
    \label{fig:geometry:2d}
  \end{subfigure}
  \caption{Test case geometry (not to scale).}
  \label{fig:geometry}
\end{figure}
\begin{table}
  \centering
  \begin{tabular}{llr}
    \toprule
    Item & Symbol & Value \\
    \midrule
    Aorta radius & $R_{\mathrm{a}}$ & \SI{1.15}{\centi\metre}\\
    Tube length & $L$ & \SI{40}{\centi\metre} \\
    Balloon length & $L_{\mathrm{b}}$ & \SI{27}{\centi\metre} \\
    Inflated radius & $R_{\max}$ & \SI{0.8}{\centi\metre}  \\
    Deflated radius & $R_{\min}$ & \SI{0.35}{\centi\metre} \\
    Catheter radius & $R_{\mathrm{c}}$ & \SI{1.25}{\milli\metre} \\
    \bottomrule
  \end{tabular}
  \caption{Geometrical dimensions of the test case.}
  \label{tab:geometry}
\end{table}
We design the test case to match the geometry of common commercial intra-aortic balloon pumps (IABPs). In particular, we use the balloon geometry of the Linear 40 cc IAB (Datascope, Fairfield, NJ, USA), which is a \SI{40}{\milli\litre} balloon with a nominal inflated diameter of \SI{1.6}{\centi\metre}. We assume axisymmetry around the central catheter axis and model the surrounding aorta as a rigid tube. Conical membrane ends with rounded edges close the balloon onto the catheter, and the deflated radius is chosen such that the stroke volume equals the nominal \SI{40}{\milli\litre}.  We use ``proximal'' to refer to the end of the tube that is closer to the heart, and ``distal'' for the end that is further away from the heart. The geometry is shown in \cref{fig:geometry} and its dimensions are summarised in \cref{tab:geometry}.

\begin{figure}
  \centering
  \newcommand\stationplot[2]{%
  \nextgroupplot[title={#2}]
  \addplot[measured, only marks, mark=*, mark size=0.5pt] table[x=t, y expr=10*\thisrow{D}] {figures/data/motion_#1_data.dat};
  \addplot[model, line width=0.9pt] table[x=t, y expr=10*\thisrow{D}] {figures/data/motion_#1_model.dat};
}
\newcommand\snapcurves[1]{%
  \addplot[primary!35, line width=0.9pt] table[x=z, y=r1] {figures/data/#1};
  \addplot[primary!51, line width=0.9pt] table[x=z, y=r2] {figures/data/#1};
  \addplot[primary!67, line width=0.9pt] table[x=z, y=r3] {figures/data/#1};
  \addplot[primary!84, line width=0.9pt] table[x=z, y=r4] {figures/data/#1};
  \addplot[primary, line width=0.9pt] table[x=z, y=r5] {figures/data/#1};
}
\newcommand\snapframe{%
  \draw[black-25, densely dotted] (axis cs:8.5, 0) -- (axis cs:8.5, 12) (axis cs:20, 0) -- (axis cs:20, 12)
    (axis cs:31.5, 0) -- (axis cs:31.5, 12);
  \node[font=\tiny, black-75, anchor=south, inner sep=1pt] at (axis cs:8.5, 12) {tip};
  \node[font=\tiny, black-75, anchor=south, inner sep=1pt] at (axis cs:20, 12) {centre};
  \node[font=\tiny, black-75, anchor=south, inner sep=1pt] at (axis cs:31.5, 12) {base};
  \draw[black-50, line width=1.2pt] (axis cs:6.0, 1.25) -- (axis cs:6.6, 1.25) (axis cs:33.4, 1.25) -- (axis cs:35, 1.25);
}
\begin{subfigure}[t]{0.32\linewidth}
  \centering
  \begin{tikzpicture}
    \begin{groupplot}[
        group style={group size=1 by 3, vertical sep=8mm, xlabels at=edge bottom, xticklabels at=edge bottom},
        clean, width=0.72\linewidth, height=1.45cm, scale only axis,
        xmin=0, xmax=0.923, ymin=5, ymax=17, xtick={0,0.4,0.8}, ytick={6,16},
        ylabel={$D$ (mm)}, xlabel={$t$ (s)},
        title style={anchor=south west, at={(0,1)}, font=\scriptsize, black-75, yshift=-1ex},
      ]
      \stationplot{tip}{tip}
      \stationplot{centre}{centre}
      \stationplot{base}{base}
    \end{groupplot}
  \end{tikzpicture}\\[0.3ex]
  {\scriptsize\color{black-75}
    \tikz[baseline=-0.6ex]\fill[measured] (0,0) circle (1.2pt);~measured \citep{Bruti2015}\quad
    \tikz[baseline=-0.6ex]\draw[model, line width=0.9pt] (0,0) -- (0.4,0);~idealised\par}
  \caption{Measurements~\cite{Bruti2015} and idealised law \eqref{eq:Rl}.}
  \label{fig:motion:stations}
\end{subfigure}\hfill
\begin{subfigure}[t]{0.64\linewidth}
  \centering
  \begin{tikzpicture}
    \begin{groupplot}[
        group style={group size=1 by 2, vertical sep=11mm, xlabels at=edge bottom, xticklabels at=edge bottom},
        clean, width=0.88\linewidth, height=2.3cm, scale only axis, clip=false,
        xmin=5, xmax=35, ymin=0, ymax=12, xtick={10,20,30}, ytick={0,4,8},
        ylabel={$\Rb$ (mm)}, xlabel={$z$ (cm)},
        title style={anchor=south west, at={(0,1)}, font=\scriptsize, black-75, yshift=1.2ex},
      ]
      \nextgroupplot[title={inflation}]
      \snapframe
      \snapcurves{snap_infl.dat}
      \node[font=\tiny, black-75] at (axis cs:30.5, 9.4) {0.08\,s};
      \node[font=\tiny, black-75] at (axis cs:25.9, 9.4) {0.11\,s};
      \node[font=\tiny, black-75] at (axis cs:19.6, 9.4) {0.14\,s};
      \node[font=\tiny, black-75] at (axis cs:13.2, 9.4) {0.17\,s};
      \node[font=\tiny, black-75] at (axis cs:9.5, 9.4) {0.20\,s};

      \nextgroupplot[title={deflation}]
      \snapframe
      \snapcurves{snap_defl.dat}
      \node[font=\tiny, black-75] at (axis cs:20, 8.7) {0.54\,s};
      \node[font=\tiny, black-75] at (axis cs:20, 2.7) {0.70\,s};
    \end{groupplot}
  \end{tikzpicture}
  \caption{Balloon radius $\Rb(z,t)$ at different instants of inflation and deflation.}
  \label{fig:motion:profiles}
\end{subfigure}
  \caption{Prescribed balloon motion profile during inflation and deflation based on measurements of~\citet{Bruti2015}}
  \label{fig:motion}
\end{figure}
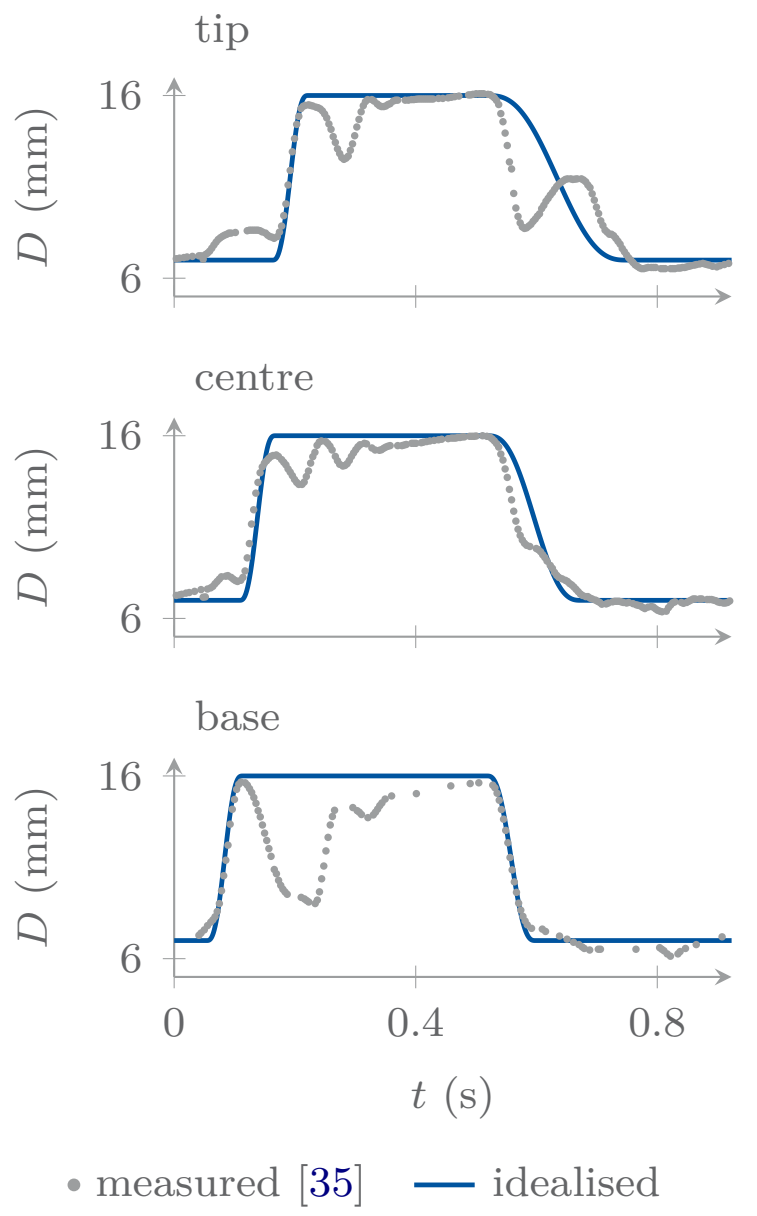
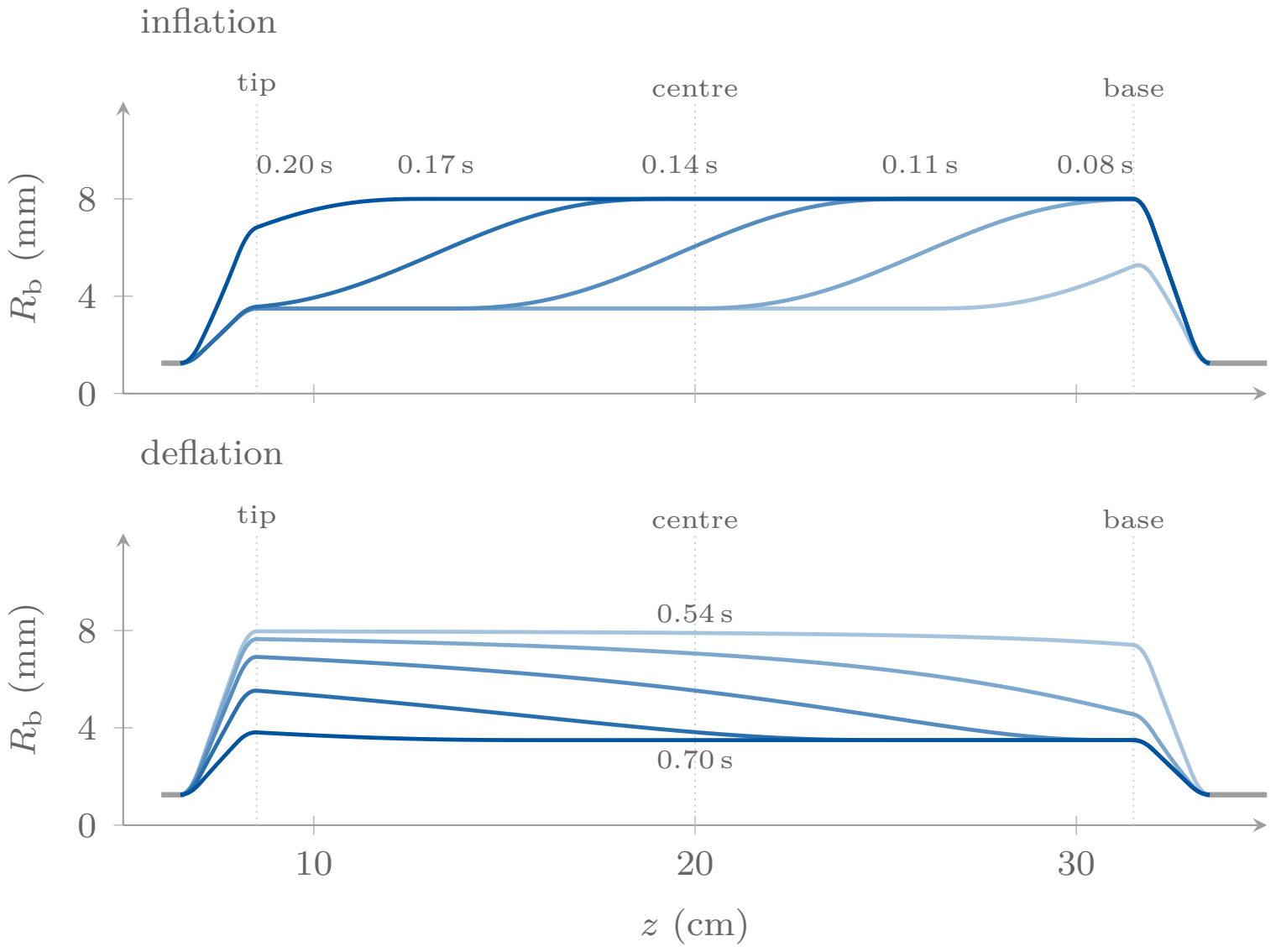

We impose the balloon motion based on data. For this purpose, we use measurements of the balloon diameter at three points along the balloon axis (tip, centre and base) from \citet{Bruti2015}. The measurements are scaled to the nominal inflated diameter of \SI{1.6}{\centi\metre} and used to construct an idealised, analytical law, see~\cref{fig:motion}. 
Overall, the balloon motion is characterised by a smooth, time-dependent deformation. Inflation is characterised by a front that travels from the base to the tip of the balloon. Deflation progresses at different rates throughout the balloon, with the base deflating fastest. Between the deflated and the inflated state, the balloon radius changes by a factor of 2.3, and the gap between the membrane and the aortic wall narrows from \SI{8.0}{\milli\metre} to \SI{3.5}{\milli\metre}. The membrane moves radially at up to \SI{14}{\centi\metre\per\second} during inflation and \SI{11}{\centi\metre\per\second} during deflation. 
The analytical formulation of balloon motion is given in \cref{app:motion}. We highlight that such analytical formulation is not necessary for the kinematic boundary conditions, as the space-time normal can be computed directly from the mesh. However, the analytical formulation is useful in the present study for generating the space-time mesh and for post-processing.

\clearpage
\subsection{Boundary conditions}
\label{sec:bc}

\begin{figure}
  \centering
  \input{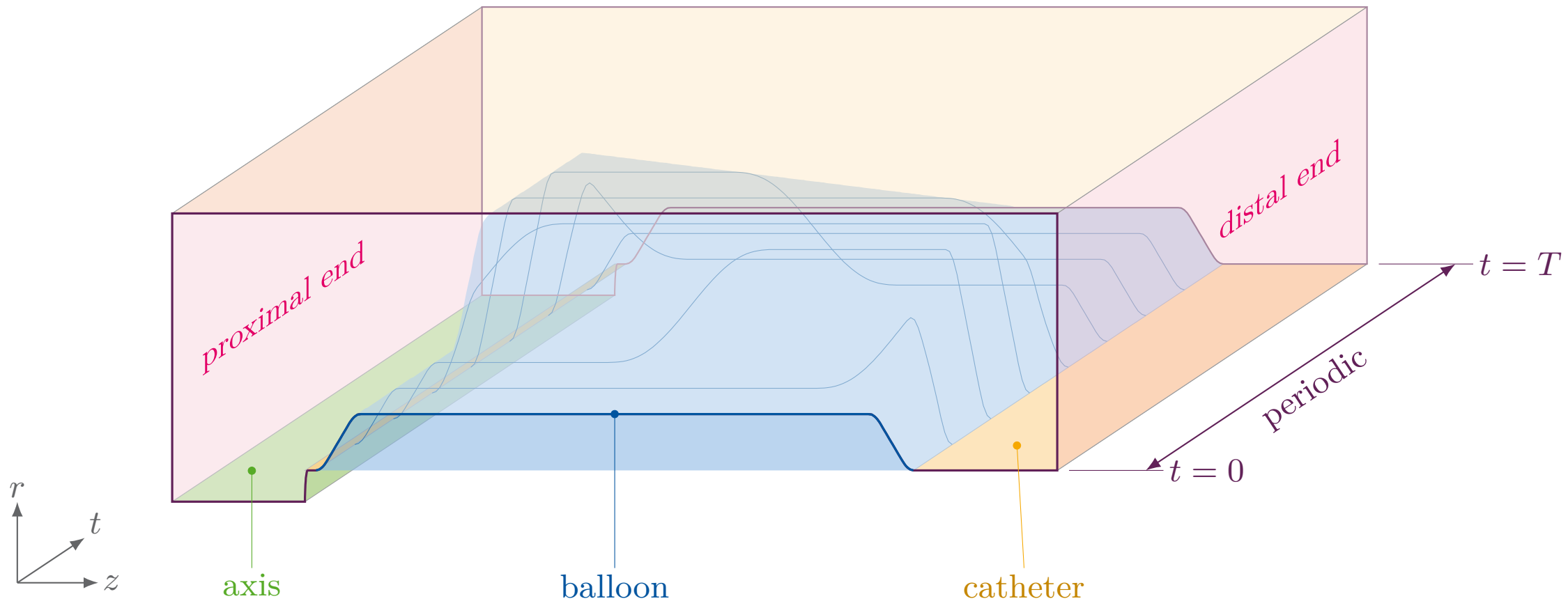}
  \caption{Space-time domain $\Q$ of the IABP test case, with boundary segments highlighted.}
  \label{fig:spacetime}
\end{figure}
The motion of the balloon is integrated into the space-time domain $\Q$. In combination with the two-dimensional axisymmetric geometry, this results in a three-dimensional space-time domain as shown in \cref{fig:spacetime}. 
The outer wall and the catheter are modelled as rigid walls with no-slip boundary conditions. The balloon is modelled as a moving wall with the kinematic boundary condition~\eqref{eq:vn} in the normal direction and no-slip in the tangential direction. The axis of symmetry is modelled with a symmetry boundary condition $u_r = 0$. This is required to avoid singularities at $r=0$. 
At the initial and final time $t=0$ and $t=T$, we impose periodic boundary conditions in time, i.e., $\uu(0)=\uu(T)$ and $p(0)=p(T)$. This is required to model the periodic nature of the flow. Due to the space-time formulation, periodicity in time can be imposed analogously to periodicity in space. This allows us to solve for the periodic steady state directly, without the need for multiple cycles.
Finally, the proximal and distal ends of the tube $\Pst_{\mathrm{ends}} = \Pst_{\mathrm{prox}} \cup \Pst_{\mathrm{dist}}$ are modelled as open boundaries:
\begin{align}
  u_r &= 0 \, , \qquad &&\text{on } \Pst_{\mathrm{ends}} \, , \label{eq:open_normal} \\
  \bm \sigma \cdot \nn &= - p(t) \nn \, , \qquad &&\text{on } \Pst_{\mathrm{ends}} \, , \label{eq:open_neumann} \\
  p(t) &= 0 \, , \qquad &&\text{on } \Pst_{\mathrm{dist}} \, , \label{eq:open_distal} \\
  p(t) &= \bar{p} + I \frac{\dd Q_{\mathrm{sys}}}{\dd t}(t) \, , \qquad &&\text{on } \Pst_{\mathrm{prox}} \, , \label{eq:open_proximal} \\
  Q_{\mathrm{sys}}(t) &= \hat Q \sin^2\!\Big(\pi\,\frac{t-t_{\mathrm{sys}}}{T_{\mathrm{sys}}}\Big) \, , \qquad
  I = 1.1 \frac{\rho L}{\pi R_{\mathrm{a}}^2} \, , \qquad &&\text{for } t_{\mathrm{sys}} \le t \le t_{\mathrm{sys}} + T_{\mathrm{sys}} \, . \label{eq:open_flow}
\end{align}
The radial velocity is set to zero at the open boundaries~\eqref{eq:open_normal} to avoid complex flow patterns at the open boundaries. The Neumann boundary condition~\eqref{eq:open_neumann} specifies a time-dependent pressure, allowing the flow to enter and leave the domain freely. This pressure is set to zero at the distal end~\eqref{eq:open_distal}, as the absolute pressure is only defined up to a constant in incompressible flow. The flow is thus driven by the time-dependent pressure at the proximal end~\eqref{eq:open_proximal}. It is composed of a constant mean pressure $\bar{p}$ and a time-dependent component. The mean pressure $\bar{p} = \SI{79}{\pascal}$ (\SI{0.59}{\mmHg}) sets the net flow over the entire cycle, which converges to about \SI{35}{\milli\litre} per cycle on the finest levels (\cref{sec:results:convergence}).
The time-dependent component models the cardiac ejection into the descending aorta during systole. The concrete form and the value of the coefficient $I$ in \cref{eq:open_flow} is based on a one-dimensional model~\cite{fryMeasurementPulsatileBlood1959}, neglecting viscous effects. The flow rate $Q_{\mathrm{sys}}$ in \cref{eq:open_flow} is modelled as a pulse beginning at $t_{\mathrm{sys}} = \SI{0.668}{\second}$ with a duration of $T_{\mathrm{sys}} = \SI{0.3}{\second}$ and a peak flow rate of $\hat Q = \SI{300}{\milli\litre\per\second}$.
In addition, we employ backflow stabilisation~\eqref{eq:backflow} at the open boundaries to avoid numerical instabilities in case of backflow.
We model blood as an incompressible Newtonian fluid with density $\rho = \SI{1060}{\kilo\gram\per\cubic\metre}$ and dynamic viscosity $\mu = \SI{3.5}{\milli\pascal\second}$. The Newtonian assumption is common for flow in large arteries, where shear rates are high.

\subsection{Space-time mesh generation}
\label{sec:mesh}
\begin{figure}
  \centering
  \input{figures/data/st_positions.tex}
\pgfplotsset{colormap={stradius}{rgb255=(255,255,255) rgb255=(199,221,242) rgb255=(64,127,183)}}
\newcommand\stlayers[1]{%
  \addplot graphics[xmin=0, xmax=40, ymin=0, ymax=0.923] {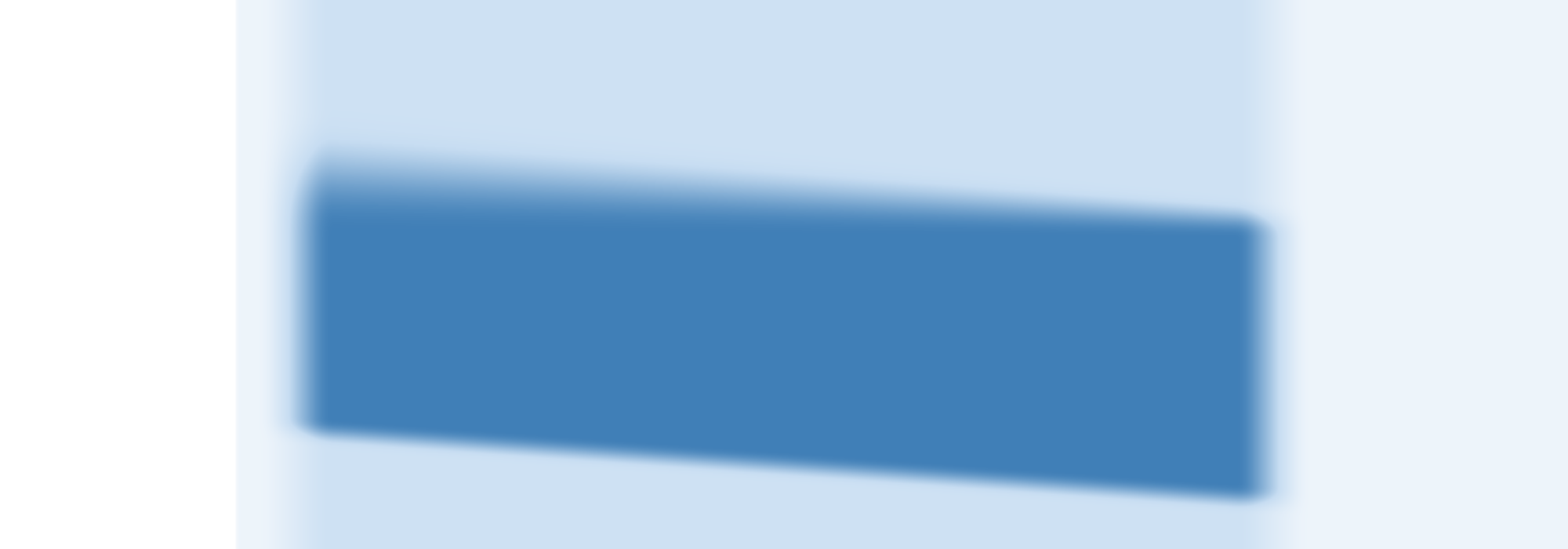};
  \addplot[black-75, line width=#1, unbounded coords=jump] table[x=z, y=t] {figures/data/st_connectors.dat};
  \addplot[black, line width=0.8pt, unbounded coords=jump] table[x=z, y=t] {figures/data/st_kinks.dat};
}
\newcommand\stslices[1]{%
  \ifnum#1=1
  \foreach \z in \ststationz {
    \edef\temp{\noexpand\draw[black-50, densely dotted, line width=0.3pt] (axis cs:\z, 0) -- (axis cs:\z, 0.923);}\temp
  }
  \fi
  \foreach \z in \stslicez {
    \edef\temp{\noexpand\draw[black, line width=0.8pt] (axis cs:\z, 0) -- (axis cs:\z, 0.923);}\temp
  }
}
\begin{subfigure}[t]{0.6\linewidth}
  \centering
  \begin{tikzpicture}
    \begin{axis}[
        clean, axis line style={black-50, -}, axis on top, enlargelimits=false,
        width=0.82\linewidth, height=5.2cm, scale only axis, clip=false,
        xmin=0, xmax=40, ymin=0, ymax=0.923, xtick={0,10,20,30,40}, ytick={0,0.2,0.4,0.6,0.8,0.923},
        yticklabels={0,0.2,0.4,0.6,0.8,$T$},
        xlabel={$z$ (cm)}, ylabel={$t$ (s)},
        colormap name=stradius, point meta min=0, point meta max=8,
        colorbar, colorbar style={
          width=2.5mm, axis lines=box, axis line style={black-50, -}, tick style={black-50},
          yticklabel pos=right, ytick={0,4,8}, ylabel={inner radius (mm)}, ylabel near ticks,
          tick label style={font=\scriptsize, color=black-75}, label style={font=\scriptsize, color=black-75},
        },
      ]
      \stlayers{0.3pt}
      \stslices{0}
      \begin{scope}[font=\scriptsize, black-75]
        \node[anchor=south] at (axis cs:7.3, 0.94) {tip};
        \node[anchor=south] at (axis cs:20, 0.94) {balloon body};
        \node[anchor=south] at (axis cs:32.5, 0.94) {base};
      \end{scope}
      \begin{scope}[font=\scriptsize]
        \node[black-75] at (axis cs:20, 0.06) {deflated};
        \node[white, rotate=-6.5] at (axis cs:15, 0.235) {inflation};
        \node[white] at (axis cs:20, 0.40) {inflated};
        \node[white] at (axis cs:13, 0.60) {deflation};
        \node[black-75] at (axis cs:20, 0.84) {deflated};
      \end{scope}
    \end{axis}
  \end{tikzpicture}
\end{subfigure}
  \caption{Topology of the inner space-time boundary in the $(z,t)$ plane}
  \label{ns:fig:topology}
\end{figure}

We generate the three-dimensional space-time domain $\Q \subset \mathbb{R}^3$ from \cref{fig:spacetime} in \textsc{Gmsh}~\cite{geuzaineGmsh3DFinite2009} using the Python API. In order to incorporate the non-uniform balloon motion, we set up a specific topology of the inner space-time boundary. This is illustrated in \cref{ns:fig:topology}, where the inner boundary is shown in the $(z,t)$ plane. This topology allows us to resolve the balloon motion continuously in time and space, while keeping the number of surfaces in \textsc{Gmsh} manageable.

We then use the built-in tetrahedral mesh generator to create an unstructured simplex mesh of the space-time domain. 
We enforce identical two-dimensional meshes on the $t = 0$ and $t = T$ surfaces and store information about the periodicity of the mesh. This is used to impose periodic boundary conditions in time. 
Inside the space-time domain, the mesh generator generally creates isotropic tetrahedra. However, the time dimension units are independent of the spatial units. Additionally, the radial dimension is much smaller than the axial dimension. Using isotropic tetrahedra would thus require a large number of elements to resolve all features. To avoid this, we scale the dimensions of the space-time domain as follows:
\begin{equation}
  \tilde{t} = \frac{t}{t_h} \, , \qquad
  \tilde{r} = \frac{r}{r_h} \, , \qquad
  \tilde{z} = z \, , \qquad
  \tilde{\Q} = \{ (\tilde{z}, \tilde{r}, \tilde{t}) \mid (z, r, t) \in \Q \} \, .
\end{equation}
We thus generate an isotropic mesh in the scaled space-time domain $\tilde{\Q}$ and then transform the mesh back to the original space-time domain $\Q$ by applying the inverse scaling. The scaling factors $t_h$ and $r_h$ determine the anisotropy of the tetrahedra in the physical space-time domain. We set $r_h = 1/8$ such that the tetrahedra are eight times finer in $r$ than in $z$. The overall resolution is controlled by the mesh size factor $f_h$, which scales the isotropic bulk element size. The mesh is refined in the channel between the balloon and the aortic wall to obtain the same number of elements across the channel throughout the balloon motion.

\begin{table}
  \centering
  \begin{tabular}{@{}l rr rrrr@{}}
    \toprule
    & \multicolumn{2}{c}{Input parameters} & \multicolumn{4}{c}{Resulting mesh} \\
    \cmidrule(lr){2-3} \cmidrule(l){4-7}
    Level & $t_h$ (\si{\milli\second\per\centi\metre}) & $f_h$ & $h_t$ (\si{\milli\second}) & $h_z$ (\si{\milli\metre}) & $h_r$ (\si{\milli\metre}) & Nodes \\
    \midrule
    B  & 23.1 & 50.0 & 8.0 & 3.46 & 0.442 & \num{134987} \\[1ex]
    T2 & 11.5 & 50.0 & 4.0 & 3.45 & 0.442 & \num{265300} \\
    T3 & 5.77 & 50.0 & 2.0 & 3.45 & 0.442 & \num{525418} \\
    T4 & 2.88 & 50.0 & 1.0 & 3.44 & 0.441 & \num{1046577} \\
    T5 & 1.44 & 50.0 & 0.5 & 3.44 & 0.442 & \num{2088296} \\[1ex]
    S2 & 32.6 & 35.4 & 8.0 & 2.46 & 0.312 & \num{253787} \\
    S3 & 46.2 & 25.0 & 8.0 & 1.75 & 0.221 & \num{483416} \\
    S4 & 65.3 & 17.7 & 8.0 & 1.24 & 0.156 & \num{935296} \\
    S5 & 92.3 & 12.5 & 8.0 & 0.88 & 0.110 & \num{1824867} \\[1ex]
    ST2 & 16.3 & 35.4 & 4.0 & 2.45 & 0.312 & \num{497442} \\
    ST3 & 11.5 & 25.0 & 2.0 & 1.74 & 0.221 & \num{1882453} \\
    ST4 & 8.16 & 17.7 & 1.0 & 1.24 & 0.156 & \num{7258527} \\
    \bottomrule
  \end{tabular}
  \caption{Hierarchy of investigated space-time meshes.}
  \label{ns:tab:levels}
\end{table}

Starting from the base level B, we generate three series of meshes, summarised in \cref{ns:tab:levels}. In the temporal series T2 to T5, we halve $t_h$ per level, which halves the mean temporal extent of the tetrahedra $h_t$ at a fixed spatial resolution. In the spatial series S2 to S5, we decrease $f_h$ by $\sqrt{2}$ per level, so that the element sizes $h_r$ and $h_z$ decrease by $\sqrt{2}$ each. We increase $t_h$ by the same factor to keep $h_t$ fixed. In the space-time series ST2 to ST4, each level combines the temporal resolution of level T$k$ with the spatial resolution of level S$k$. Thus, $h_t$ is halved while $h_z$ and $h_r$ decrease by $\sqrt{2}$ each. In the temporal and spatial series, the number of nodes roughly doubles from one level to the next. In the space-time series, it roughly quadruples.

\subsection{Numerical scheme}
\label{sec:partitioning}

We use the multiphysics finite element solver \texttt{XNS}~\cite{dirkesEulerianFormulationTensorbased2024} to solve the discrete problem. This consists of two steps. First, the nonlinear system of equations is solved using a Newton-Raphson method with Armijo line search. We solve up to a relative residual tolerance of $10^{-8}$.
We solve the linearised systems with the restarted GMRES method~\cite{y.saadGMRESGeneralizedMinimal1986} with right ILUT preconditioning~\cite{saadILUTDualThreshold1994}, a Krylov space of dimension 300, a relative tolerance of $10^{-6}$ and at most 6000 iterations per Newton step.

\begin{figure}
  \centering
  \input{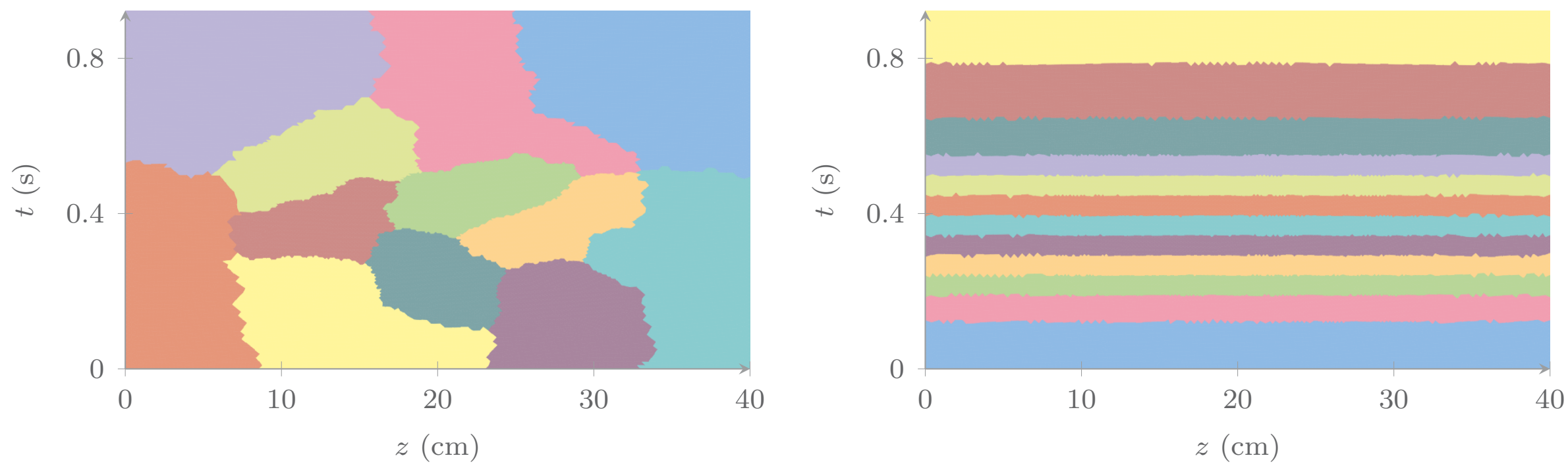}
  \caption{Decomposition of the space-time mesh (level B) into 12 subdomains at $r = R_a$.}
  \label{fig:partitioning}
\end{figure}

To ensure parallel scalability, the space-time mesh is decomposed into subdomains, which are distributed over the available MPI ranks. The standard approach is to minimise the interface between subdomains, thus reducing the number of interface nodes, which we do using METIS. This results in compact subdomains that cut through space and time alike, as shown in~\cref{fig:partitioning:metis}. We refer to this as mixed partitioning.
An alternative approach is to perform the decomposition primarily in time, as illustrated in~\cref{fig:partitioning:slabs}. This creates larger interfaces, but allows for a more natural decomposition of the problem for parallelism in time. In this case, each subdomain contains the entire spatial domain.
The preconditioner acts on each subdomain separately, neglecting the coupling between subdomains. The stronger this coupling, the less effective the preconditioner, and the more GMRES iterations are needed. The choice of partitioning therefore affects convergence of GMRES. This effect is investigated in \cref{sec:results:partitioning}.

\section{Results}
\label{sec:results}

In this section, we present the results of the IABP test case. We first analyse the resulting flow field, including flow rates, velocity snapshots and wall shear stress. We then study convergence under temporal, spatial and space-time refinement. Finally, we present measurements of solver performance and weak scaling.

\subsection{Flow field}
\label{sec:results:flow}
\begin{figure}
  \centering
  \input{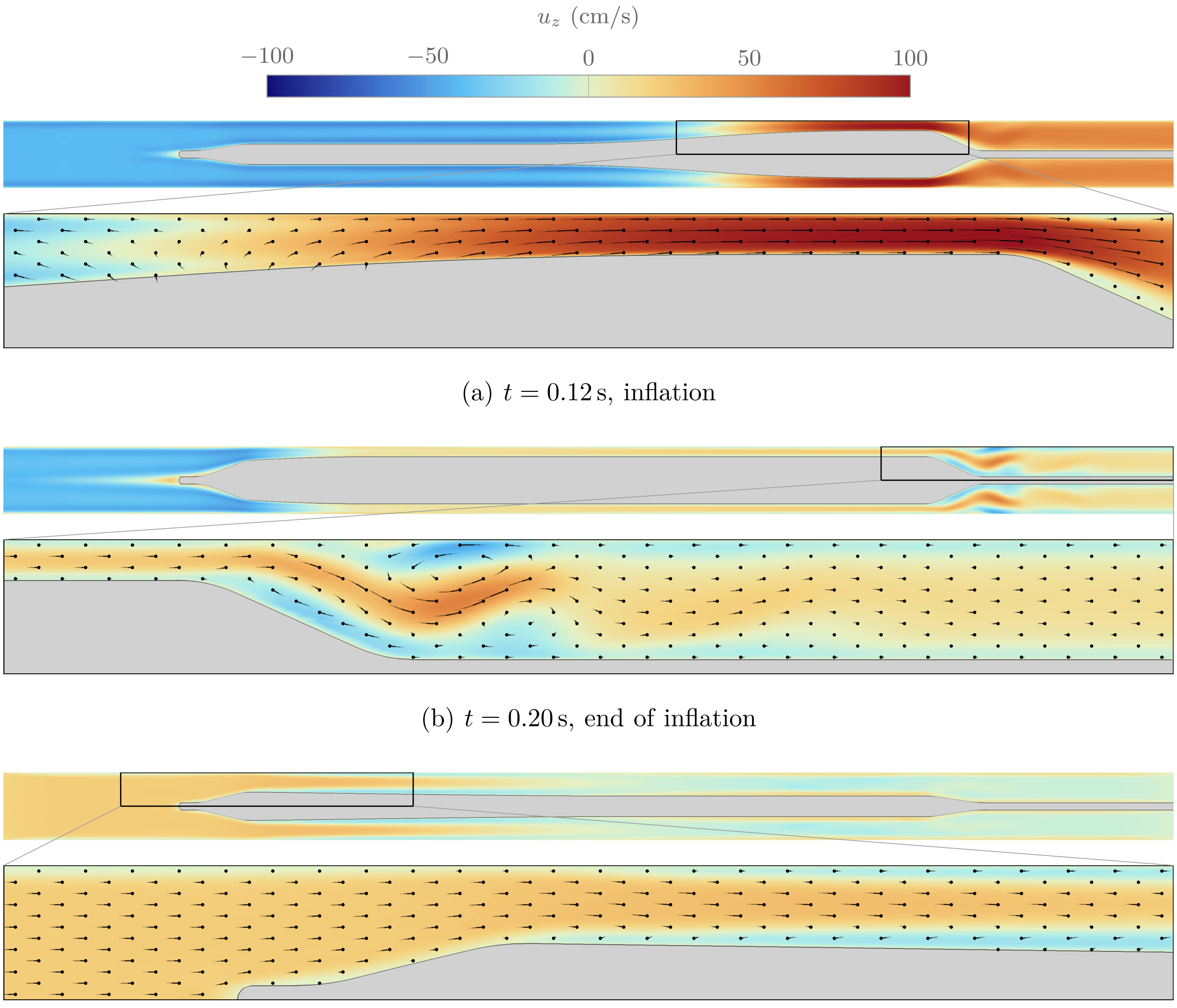}
\pgfmathsetlengthmacro\flowW{\linewidth}
\pgfmathsetlengthmacro\flowHov{\linewidth * 2.3 / 40}
\pgfmathsetlengthmacro\flowHzm{\linewidth * 1.15 / 10}
\pgfmathsetlengthmacro\flowWcb{0.55 * \linewidth}       %
\pgfplotsset{
  flowbase/.style={
    hide axis, enlargelimits=false, scale only axis, width=\flowW,
    colormap name=pvfast, point meta min=-\flowvmax, point meta max=\flowvmax,
  },
  flowoverview/.style={flowbase, height=\flowHov, xmin=0, xmax=40, ymin=-1.15, ymax=1.15},
  flowzoom/.style={flowbase, height=\flowHzm, ymin=0, ymax=1.15},
  trail/.style={black, line cap=round, unbounded coords=jump, empty line=jump,
    restrict expr to domain={\thisrow{part}}{#1:#1}},
}
\newcommand\flowrow[6]{%
  \pgfmathsetmacro\zend{#3 + 10}
  \begin{axis}[flowoverview, name=ov#1, #6]
    \addplot graphics[xmin=0, xmax=40, ymin=-1.15, ymax=1.15] {figures/flow_closeups/overview_#2.png};
    \draw[black, line width=0.5pt] (axis cs:#3,0) rectangle (axis cs:\zend,1.15);   %
    \coordinate (ov#1l) at (axis cs:#3,0);
    \coordinate (ov#1r) at (axis cs:\zend,0);
  \end{axis}
  \begin{axis}[flowzoom, name=zm#1, at={(ov#1.south west)}, anchor=north west, yshift=-3.5mm, xmin=#3, xmax=\zend]
    \addplot graphics[xmin=#3, xmax=\zend, ymin=0, ymax=1.15] {figures/flow_closeups/zoom_#2.png};
    \addplot[trail=1, line width=0.25pt, opacity=0.25] table {figures/data/closeup_trails_#2.dat};
    \addplot[trail=2, line width=0.35pt, opacity=0.5] table {figures/data/closeup_trails_#2.dat};
    \addplot[trail=3, line width=0.45pt, opacity=0.85] table {figures/data/closeup_trails_#2.dat};
    \addplot[only marks, mark=*, mark size=0.45pt, black] table {figures/data/closeup_heads_#2.dat};
    \draw[black, line width=0.5pt] (rel axis cs:0,0) rectangle (rel axis cs:1,1);
  \end{axis}
  \draw[black-50, line width=0.3pt] (ov#1l) -- (zm#1.north west) (ov#1r) -- (zm#1.north east);   %
  \node[name=cap#1, anchor=north, text width=\flowW, inner sep=0pt, yshift=-2mm]
    at (zm#1.south) {\subcaption{$t=\SI{#2}{\second}$, #4}\label{#5}};
}
\begin{tikzpicture}
  \flowrow{a}{0.12}{23}{inflation}{fig:flowfield:inflation}{
    colorbar horizontal, colorbar style={
      at={(0.5,1.35)}, anchor=south, width=\flowWcb, height=3mm,
      axis lines=box, axis line style={black-50, -}, tick style={black-50},
      xtick={-100,-50,0,50,100}, xticklabel pos=upper, xlabel near ticks, xlabel={$u_z$ (cm/s)},
      tick label style={font=\footnotesize, color=black-75}, label style={font=\footnotesize, color=black-75},
    }}
  \flowrow{b}{0.20}{30}{end of inflation}{fig:flowfield:front}{
    at={(zma.south west |- capa.south)}, anchor=north west, yshift=-5mm}
  \flowrow{c}{0.66}{4}{end of deflation}{fig:flowfield:deflation}{
    at={(zmb.south west |- capb.south)}, anchor=north west, yshift=-5mm}
\end{tikzpicture}
  \vspace{-1em}
  \caption{Axial velocity at three instants (ST4).}
  \label{fig:flowfield}
\end{figure}

First, we look at the flow field in the spatial domain $\Omega(t)$ at three instants of the period. This is shown in \cref{fig:flowfield}. Each instant is shown in two views: an overview of the whole tube (mirrored about the axis) and a close-up with particle trails. The colour map is the same for all three instants and shows the axial velocity $u_z$. \Cref{fig:flowfield:inflation} shows the inflation phase. The inflation front travels from the base towards the tip, displacing fluid in both directions. In particular, the location of the inflation front is visible as a band of zero axial velocity. Fluid closer to the tip is displaced towards the proximal end, while fluid closer to the base is displaced towards the distal end. \Cref{fig:flowfield:front} shows the end of inflation, when the inflation front has reached the tip. At this stage, a jet has formed at the base. As a result, there is recirculation between the catheter and the outer wall. \Cref{fig:flowfield:deflation} shows the end of deflation. In this stage, fluid is drawn in from both ends. The positive axial velocity dominates in the proximal half of the tube, coinciding with the beginning of systole. In the distal half, the flow is still reversed. A thin layer of reverse flow reaches along the walls towards the tip, which deflates last.

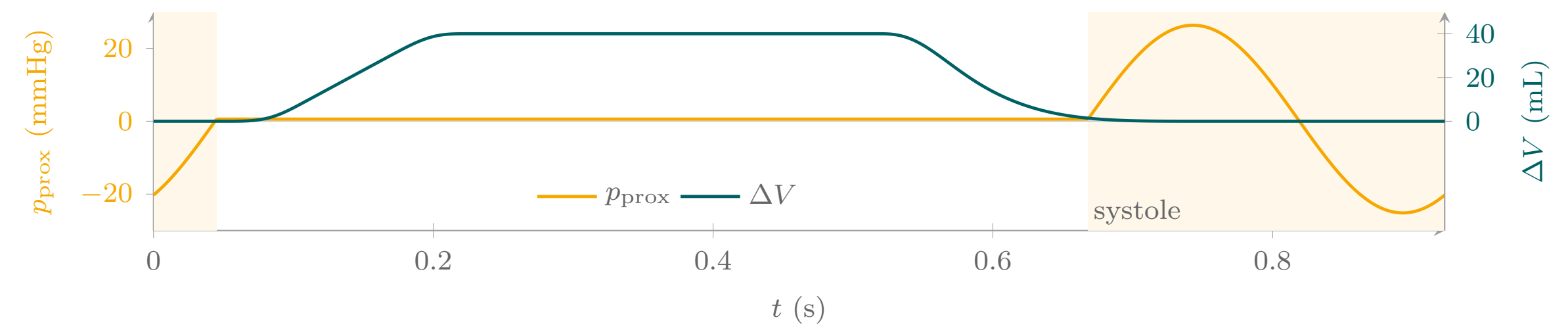
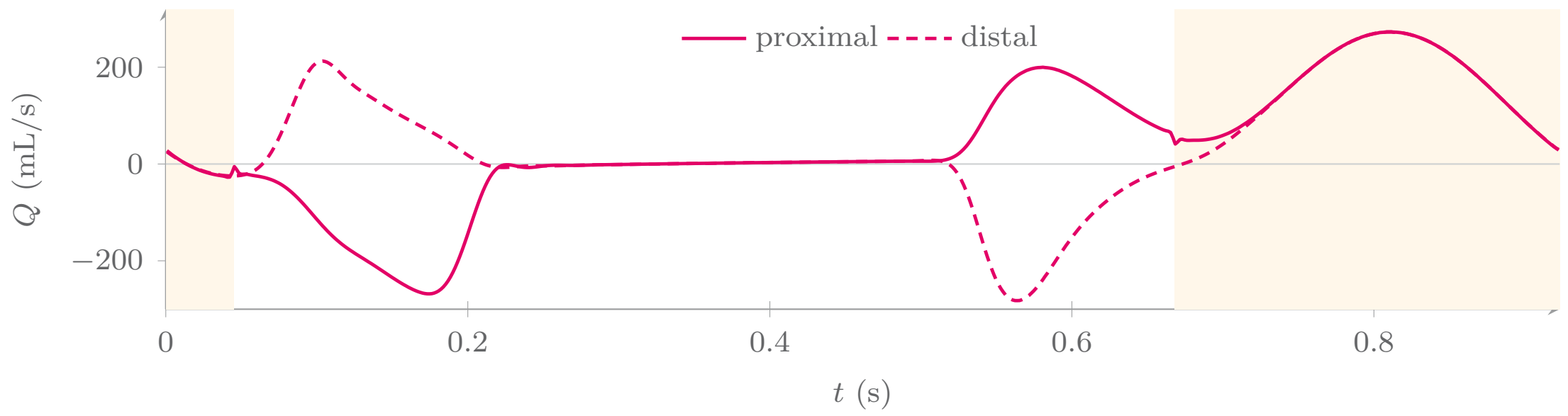
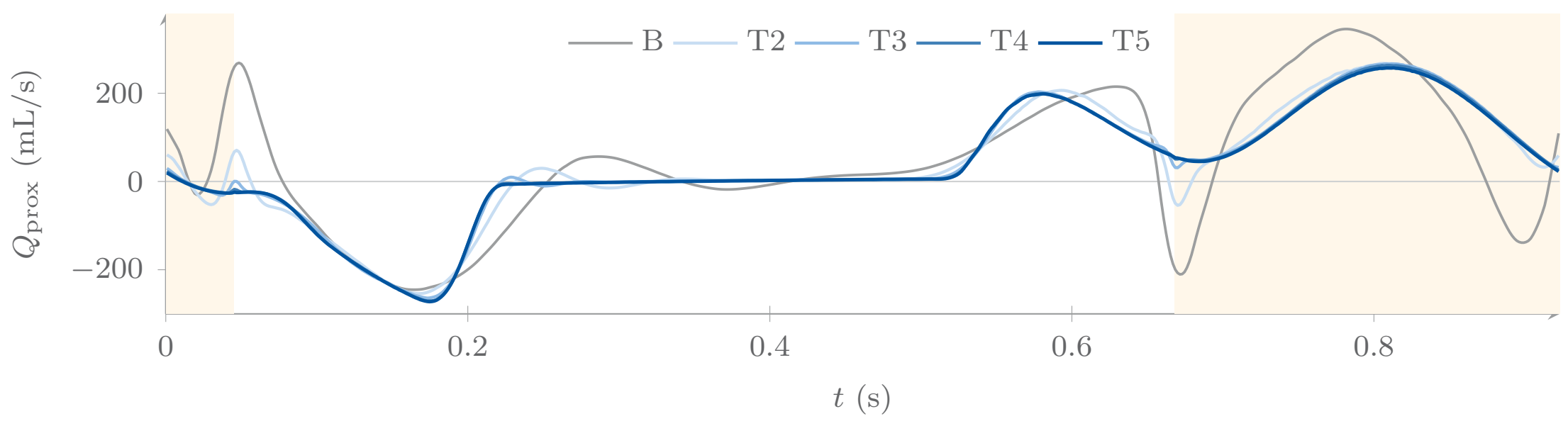
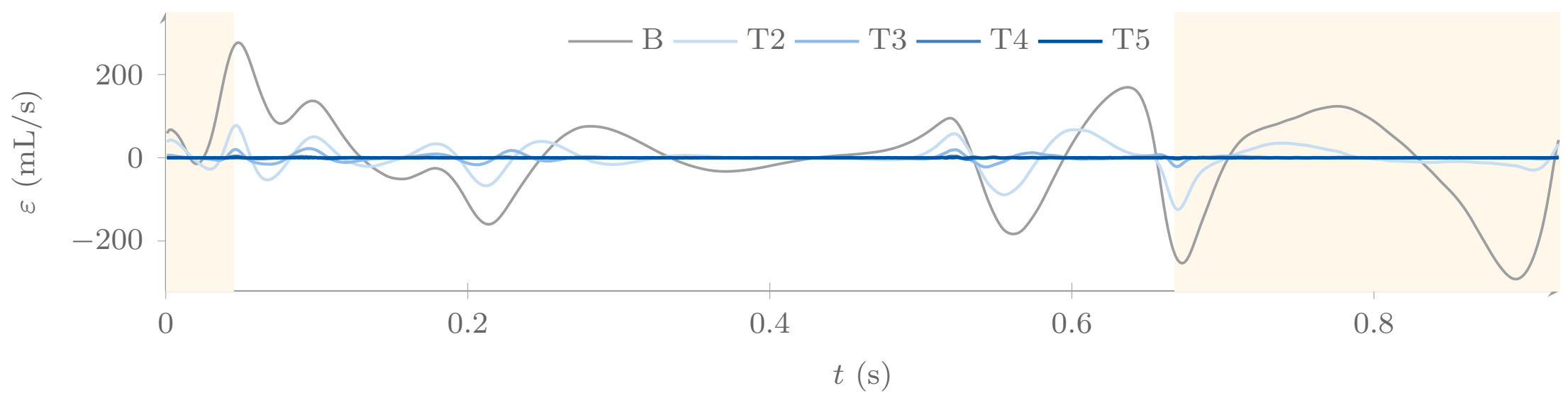
\begin{figure}
  \centering
  \tikzset{lB/.style={black-50, line width=0.7pt}, lT2/.style={primary-25, line width=0.8pt},
  lT3/.style={primary-50, line width=0.8pt}, lT4/.style={primary-75, line width=0.8pt}, lT5/.style={primary, line width=0.9pt},
  lST4/.style={magenta, line width=0.9pt}}
\pgfplotsset{
  stack/.style={clean, width=\linewidth-30mm, scale only axis, xmin=0, xmax=0.923,
    xtick={0,0.2,0.4,0.6,0.8}, xlabel={$t$ (s)}, enlarge x limits=false,
    legend style={draw=none, fill=none, font=\scriptsize, text=black-75, at={(0.5,1)}, anchor=north},
    legend cell align=left},
}
\captionsetup[subfigure]{skip=1pt}
\newcommand\sysband[2]{%
  \fill[systole] (axis cs:0.668,#1) rectangle (axis cs:0.923,#2);
  \fill[systole] (axis cs:0,#1) rectangle (axis cs:0.045,#2);
  \draw[black-25, thin] (axis cs:0,0) -- (axis cs:0.923,0);
}
\newcommand\timereflevels[2]{%
  \addplot[lB] table[x=t, y=#1] {figures/data/#2_B.dat};
  \addplot[lT2] table[x=t, y=#1] {figures/data/#2_T2.dat};
  \addplot[lT3] table[x=t, y=#1] {figures/data/#2_T3.dat};
  \addplot[lT4] table[x=t, y=#1] {figures/data/#2_T4.dat};
  \addplot[lT5] table[x=t, y=#1] {figures/data/#2_T5.dat};
}
\begin{subfigure}{\linewidth}
  \centering
  \begin{tikzpicture}[trim axis left, trim axis right]
    \begin{axis}[
        stack, height=2.2cm, ymin=-30, ymax=30, ytick={-20,0,20},
        ylabel={$p_{\mathrm{prox}}$ (mmHg)}, ylabel style={color=orange}, yticklabel style={color=orange},
      ]
      \sysband{-30}{30}
      \addplot[orange, line width=0.9pt] table[x=t, y=p] {figures/data/pressure.dat};
      \node[font=\scriptsize, text=black-75, anchor=south west, inner sep=1.5pt] at (axis cs:0.668,-30) {systole};
    \end{axis}
    \begin{axis}[
        stack, height=2.2cm, axis x line=none, xlabel={}, axis y line*=right, ymin=-50, ymax=50, ytick={0,20,40},
        ylabel={$\Delta V$ (mL)}, ylabel style={color=petrol}, yticklabel style={color=petrol},
        legend columns=2, legend style={at={(0.4,0.03)}, anchor=south},
      ]
      \addlegendimage{orange, line width=0.9pt}
      \addplot[petrol, line width=0.9pt] table[x=t, y=dV] {figures/data/flow_dthalf.dat};
      \legend{$p_{\mathrm{prox}}$, $\Delta V$}
    \end{axis}
  \end{tikzpicture}
  \caption{Prescribed proximal pressure $p_{\mathrm{prox}}$ and balloon volume $\Delta V=V-V_{\min}$.}
  \label{fig:flowrates:p}
\end{subfigure}

\vspace{6mm}
\begin{subfigure}{\linewidth}
  \centering
  \begin{tikzpicture}[trim axis left, trim axis right]
    \begin{axis}[
        stack, height=2.8cm, ymin=-300, ymax=320, ytick={-200,0,200}, ylabel={$Q$ (mL/s)}, legend columns=2,
      ]
      \sysband{-300}{320}
      \addplot[lST4] table[x=t, y=Qin] {figures/data/qflow_ST4.dat};
      \addplot[lST4, densely dashed] table[x=t, y=Qout] {figures/data/qflow_ST4.dat};
      \legend{proximal, distal}
    \end{axis}
  \end{tikzpicture}
  \caption{Flow rate through the proximal and distal end, level ST4.}
  \label{fig:flowrates:Q}
\end{subfigure}

\vspace{6mm}
\begin{subfigure}{\linewidth}
  \centering
  \begin{tikzpicture}[trim axis left, trim axis right]
    \begin{axis}[
        stack, height=2.8cm, ymin=-300, ymax=380, ytick={-200,0,200}, ylabel={$Q_{\mathrm{prox}}$ (mL/s)},
        legend columns=5,
      ]
      \sysband{-300}{380}
      \timereflevels{Qin}{qflow}
      \legend{B, T2, T3, T4, T5}
    \end{axis}
  \end{tikzpicture}
  \caption{Proximal flow rate on the temporal levels B and T2--T5.}
  \label{fig:flowrates:Qtime}
\end{subfigure}

\vspace{6mm}
\begin{subfigure}{\linewidth}
  \centering
  \begin{tikzpicture}[trim axis left, trim axis right]
    \begin{axis}[
        stack, height=2.6cm, ymin=-320, ymax=350, ytick={-200,0,200}, ylabel={$\varepsilon$ (mL/s)},
        legend columns=5,
      ]
      \sysband{-320}{350}
      \timereflevels{epsavg}{mass}
      \legend{B, T2, T3, T4, T5}
    \end{axis}
  \end{tikzpicture}
  \caption{Mass error $\varepsilon(t)$ under time refinement.}
  \label{fig:flowrates:mass}
\end{subfigure}
  \caption{Global quantities over one period.}
  \label{fig:flowrates}
\end{figure}
In \cref{fig:flowrates}, we show the global quantities over one period. \Cref{fig:flowrates:p} shows the boundary conditions: the proximal pressure according to \cref{eq:open_proximal} and the balloon volume $\Delta V = V(t)-V_0$. Systole is shaded in yellow. \Cref{fig:flowrates:Q} shows the resulting flow rates $Q(t)=\int_{\Gamma}\uu\cdot\nn\,\dd s$ at the proximal and distal ends on the finest space-time level ST4. A positive flow rate corresponds to flow in the positive $z$ direction. During inflation, there is a significant delay between the peak flow rates at the two ends. The distal end reaches its peak first, while the proximal end reaches its peak later. During deflation, this trend is reversed, but it is not as pronounced. These are effects of the non-uniform inflation behaviour of the balloon.
\Cref{fig:flowrates:Qtime} analyses the proximal flow rate under temporal refinement. On the coarsest level B, there are strong oscillations at the beginning and end of systole, as well as after inflation. These lead to non-physical negative flow rates locally. The oscillations occur at the instants where the time derivative of the prescribed proximal pressure $\dd p_{\mathrm{prox}}/\dd t$ is discontinuous, or where the flow rate changes rapidly. The oscillations are progressively reduced with temporal refinement. They are thus an artefact of continuous time discretisation in the presence of discontinuities or strong gradients, similar to Gibbs phenomena.
\Cref{fig:flowrates:mass} shows the local mass error
\begin{equation*}
  \varepsilon(t) = Q_{\mathrm{prox}}(t) - Q_{\mathrm{dist}}(t) - Q_{\mathrm{b}}(t) \, , \qquad
  Q_{\mathrm{b}}(t) = \int_{\Gamma_{\mathrm{b}}(t)} \uu\cdot\nn\,\dd s \, ,
\end{equation*}
where $Q_{\mathrm{b}}$ is the flux through the balloon membrane. The local mass balance is averaged over \SI{7.7}{\milli\second}, which corresponds to about one element width $h_t$ of the coarsest temporal level B. Analytically, $\varepsilon \equiv 0$ for a divergence-free velocity field. The coarsest level B shows large local mass errors, up to \SI{290}{\milli\litre\per\second} during systole. The local mass error is reduced with temporal refinement, down to \SI{0.2}{\milli\litre\per\second} (rms) on T5. The global mass balance
\begin{equation*}
  E = \int_0^T \varepsilon(t)\,\dd t 
\end{equation*}
is approximately zero on all levels. This means that total mass is conserved over a full cycle, but not locally in time. This is an effect of the continuous space-time formulation of the problem, which only enforces the divergence-free condition in a weak sense. The local mass error thus needs to be controlled by sufficient temporal resolution. 

\begin{figure}
  \centering
  \input{figures/data/wss_range.tex}
\pgfmathsetmacro\wssmaxPa{\wssmax/10}%
\pgfplotsset{colormap={diverging}{rgb255=(0,84,159) rgb255=(142,186,229) rgb255=(255,255,255) rgb255=(230,150,121) rgb255=(204,7,30)}}
\begin{tikzpicture}
  \begin{axis}[
      clean, width=0.75\linewidth, height=4.5cm, scale only axis, enlargelimits=false,
      xmin=0, xmax=40, ymin=0, ymax=0.923, xtick={0,10,20,30,40}, ytick={0,0.4,0.8},
      xlabel={$z$ (cm)}, ylabel={$t$ (s)}, axis on top,
      colormap name=diverging, point meta min=-\wssmaxPa, point meta max=\wssmaxPa,
      colorbar, colorbar style={width=2.5mm, ytick={-\wssmaxPa,0,\wssmaxPa},
        yticklabel style={font=\scriptsize, color=black-75}, ylabel={$\tau_w$ (\si{\pascal})},
        ylabel style={font=\scriptsize, color=black-75}},
    ]
    \addplot graphics[xmin=0, xmax=40, ymin=0, ymax=0.923] {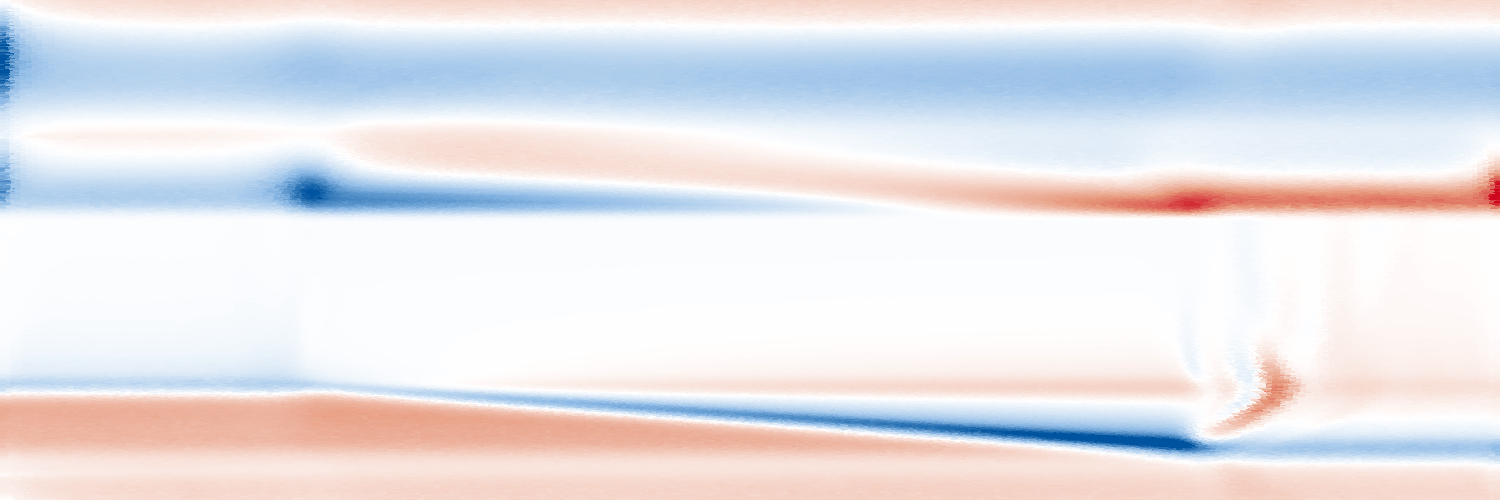};
  \end{axis}
\end{tikzpicture}%
  \caption{Wall shear stress $\tau_w(z,t)$ on the aortic wall over one period (ST4).}
  \label{fig:wss}
\end{figure}
The resulting wall shear stress $\tau_w(z,t)$ on the aortic wall is shown in \cref{fig:wss}. It shows significant magnitudes of up to about \SI{12}{\pascal} and strong temporal variations. The inflation front and the deflation front are visible as bands of high wall shear stress travelling along the balloon. The highest values occur over the balloon base and tip, where the flow is accelerated and decelerated. The wall shear stress reverses during deflation.
Such wall shear stresses can have a significant effect on the endothelial cells of the aortic wall, which are sensitive to strong oscillatory wall shear stresses. However, IABP therapy is generally applied temporarily, while remodelling of the aortic wall occurs over longer time scales~\cite{rannoComputationalModelCoronary2025}.

\subsection{Convergence behaviour}
\label{sec:results:convergence}

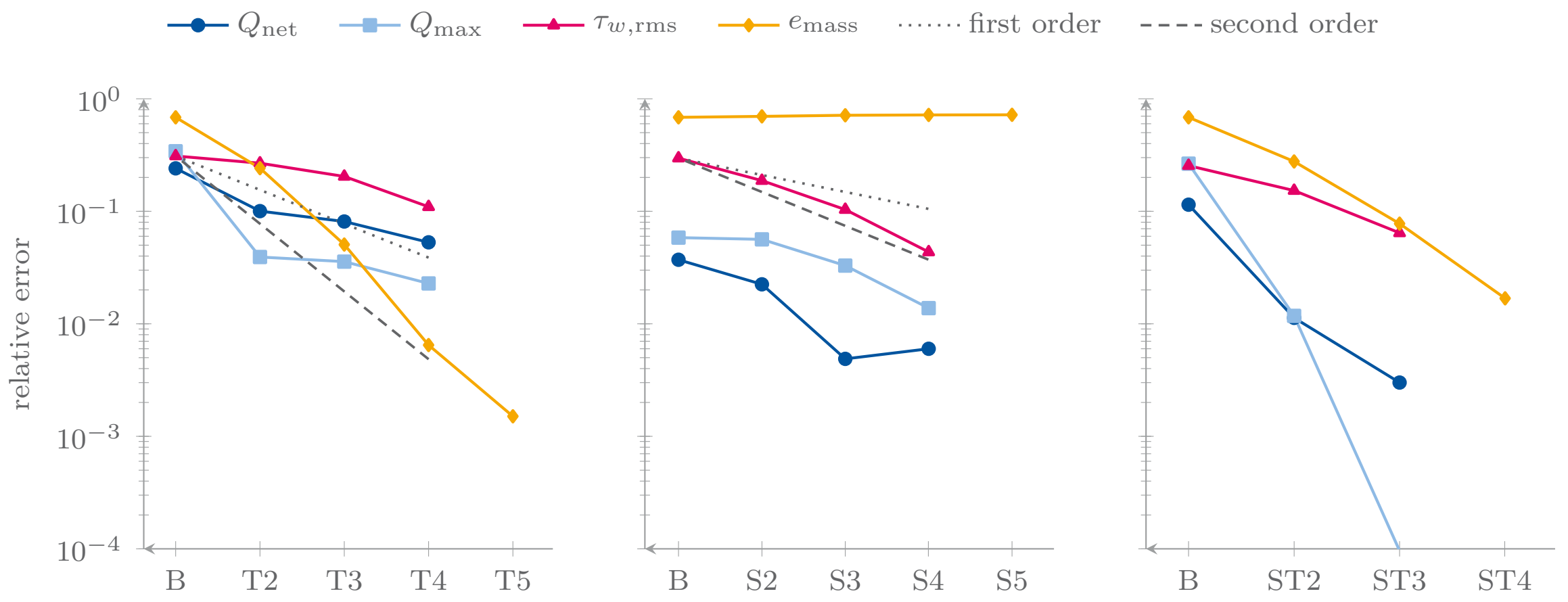
\begin{figure}
  \centering
  \tikzset{refline/.style={black-75, line width=0.7pt}}
\begin{tikzpicture}
  \begin{groupplot}[
      group style={group size=3 by 1, horizontal sep=9mm, yticklabels at=edge left, ylabels at=edge left},
      clean, width=0.25\linewidth, height=4.4cm, scale only axis, xmode=log, ymode=log,
      x dir=reverse, ymin=1e-4, ymax=1,
      ylabel={relative error},
      every axis plot/.append style={line width=0.8pt, mark size=1.6pt},
    ]
    \nextgroupplot[xmin=0.045, xmax=1.3, xtick={1,0.5,0.25,0.125,0.0625}, xticklabels={B,T2,T3,T4,T5},
        legend style={draw=none, fill=none, font=\scriptsize, text=black-75, /tikz/every even column/.append style={column sep=3mm}},
        legend columns=6, legend cell align=left, legend to name=convlegend]
    \addplot[primary, mark=*] table[x=h, y=net] {figures/data/conv_time.dat};
    \addplot[primary-50, mark=square*] table[x=h, y=qmax] {figures/data/conv_time.dat};
    \addplot[magenta, mark=triangle*] table[x=h, y=taurms] {figures/data/conv_time.dat};
    \addplot[orange, mark=diamond*] table[x=h, y=mass] {figures/data/convmass_time.dat};
    \addplot[refline, dotted, domain=0.125:1, samples=2, mark=none] {0.310*x};
    \addplot[refline, dash pattern=on 3pt off 2pt, domain=0.125:1, samples=2, mark=none] {0.310*x^2};
    \legend{$Q_\mathrm{net}$, $Q_\mathrm{max}$, $\tau_{w,\mathrm{rms}}$, $e_\mathrm{mass}$, first order, second order}
    \nextgroupplot[xmin=0.21, xmax=1.15, xtick={1,0.707,0.5,0.354,0.25}, xticklabels={B,S2,S3,S4,S5}]
    \addplot[primary, mark=*] table[x=h, y=net] {figures/data/conv_space.dat};
    \addplot[primary-50, mark=square*] table[x=h, y=qmax] {figures/data/conv_space.dat};
    \addplot[magenta, mark=triangle*] table[x=h, y=taurms] {figures/data/conv_space.dat};
    \addplot[orange, mark=diamond*] table[x=h, y=mass] {figures/data/convmass_space.dat};
    \addplot[refline, dotted, domain=0.354:1, samples=2, mark=none] {0.297*x};
    \addplot[refline, dash pattern=on 3pt off 2pt, domain=0.354:1, samples=2, mark=none] {0.297*x^2};
    \nextgroupplot[xmin=0.3, xmax=1.15, xtick={1,0.707,0.5,0.354}, xticklabels={B,ST2,ST3,ST4}]
    \addplot[primary, mark=*] table[x=h, y=net] {figures/data/conv_spacetime.dat};
    \addplot[primary-50, mark=square*] table[x=h, y=qmax] {figures/data/conv_spacetime.dat};
    \addplot[magenta, mark=triangle*] table[x=h, y=taurms] {figures/data/conv_spacetime.dat};
    \addplot[orange, mark=diamond*] table[x=h, y=mass] {figures/data/convmass_spacetime.dat};
  \end{groupplot}
  \coordinate (capy) at (current bounding box.south);
  \node[anchor=south, inner sep=0pt, yshift=2mm] at (current bounding box.north) {\pgfplotslegendfromname{convlegend}};
  \panelcaption{group c1r1.south |- capy}{0.3\linewidth}{Temporal refinement.}{fig:convergence:time}
  \panelcaption{group c2r1.south |- capy}{0.3\linewidth}{Spatial refinement.}{fig:convergence:space}
  \panelcaption{group c3r1.south |- capy}{0.3\linewidth}{Space-time refinement.}{fig:convergence:spacetime}
\end{tikzpicture}
  \caption{Convergence of the continuous space-time approach using the levels shown in \cref{ns:tab:levels}.}
  \label{fig:convergence}
\end{figure}
We study convergence of the continuous space-time approach by four quantities:
\begin{align}
  \text{net flow per period: } \qquad & Q_\mathrm{net} = \int_0^T Q_{\mathrm{prox}}(t)\,\dd t \, , \label{eq:convergence:net} \\
  \text{peak proximal flow rate: } \qquad & Q_{\mathrm{max}} = \max_{t\in[0,T]} Q_{\mathrm{prox}}(t) \, , \label{eq:convergence:max} \\
  \text{aortic wall shear stress: } \qquad & \tau_{w,\mathrm{rms}}
  = \sqrt{ \frac{1}{TL} \int_0^T \int_{0}^{L} \tau_w^2(z,t)\,\dd z\,\dd t } \, ,
  \label{eq:convergence:tau} \\
  \text{relative mass error: } \qquad & e_{\mathrm{mass}} = \frac{\|\varepsilon\|_{L^2(0,T)}}{\|Q_{\mathrm{prox}}\|_{L^2(0,T)}} 
  = \sqrt{\frac{{\int_0^T \varepsilon^2(t)\,\dd t}}{{\int_0^T Q_{\mathrm{prox}}^2(t)\,\dd t}}}
  \, . 
  \label{eq:convergence:mass}
\end{align}
The results are shown in \cref{fig:convergence}. 
For the quantities \labelcref{eq:convergence:net,eq:convergence:max,eq:convergence:tau}, no exact solution is available, so we use the finest level of each series as a reference. We calculate the relative error $|f_k - f_{\mathrm{ref}}|/|f_{\mathrm{ref}}|$ of level $k$ to the finest level of each series. This can only be done up to the second finest level. The relative mass error~\labelcref{eq:convergence:mass} can be evaluated on all levels, since the exact value is zero. 

The temporal series in \cref{fig:convergence:time} shows that the net flow, peak flow rate and wall shear stress converge roughly with first order in time. These quantities are dominated by the spatial discretisation. The relative mass error, on the other hand, converges with second order in time. This shows that the local mass error is dominated by the temporal discretisation, which is consistent with the results in \cref{fig:flowrates:mass}. The spatial series in \cref{fig:convergence:space} shows that the mass error is not reduced by spatial refinement, while the other quantities converge roughly with first to second order in space. The net flow $Q_\mathrm{net}$ stagnates beyond level S3, as the temporal error starts to dominate. 
The space-time series in \cref{fig:convergence:spacetime} does not have reference slopes, since they would mix the orders of the spatial and temporal discretisation. As expected, all quantities converge under space-time refinement, as this combines the effects of temporal and spatial refinement. The relative mass error is reduced to \SI{2}{\percent} on the finest level ST4.

\subsection{Solver performance and weak scaling}
\label{sec:results:scaling}
\label{sec:results:partitioning}

\begin{figure}[htbp]
  \centering
  \begin{tikzpicture}
  \begin{groupplot}[
      group style={group size=3 by 1, horizontal sep=25mm},
      clean, width=0.2\linewidth, height=3.6cm, scale only axis,
      xmode=log, log basis x=2, xmin=45, xmax=1300, xtick={60,240,960}, xticklabels={60,240,960},
      xlabel={ranks}, ymin=0,
      every axis plot/.append style={line width=0.8pt, mark=*, mark size=1.6pt},
    ]
    \nextgroupplot[ylabel={time per GMRES it.\ (ms)}, ymax=70,
        legend style={draw=none, fill=none, font=\scriptsize, text=black-75}, legend columns=3,
        legend to name=weaklegend]
    \addplot[primary] table[x=ranks, y=msit] {figures/data/weak_T.dat};
    \addplot[magenta] table[x=ranks, y=msit] {figures/data/weak_S.dat};
    \addplot[grun] table[x=ranks, y=msit] {figures/data/weak_ST.dat};
    \legend{temporal (T), spatial (S), space-time (ST)}
    \nextgroupplot[ylabel={assembly per Newton step (s)}, ymax=3]
    \addplot[primary] table[x=ranks, y=asm] {figures/data/weak_T.dat};
    \addplot[magenta] table[x=ranks, y=asm] {figures/data/weak_S.dat};
    \addplot[grun] table[x=ranks, y=asm] {figures/data/weak_ST.dat};
    \nextgroupplot[ylabel={GMRES its.\ per Newton step}, ymax=5000, scaled y ticks=false,
        ytick={0,1000,2000,3000,4000,5000}]
    \addplot[primary] table[x=ranks, y=gmres] {figures/data/weak_T.dat};
    \addplot[magenta] table[x=ranks, y=gmres] {figures/data/weak_S.dat};
    \addplot[grun] table[x=ranks, y=gmres] {figures/data/weak_ST.dat};
  \end{groupplot}
  \coordinate (capy) at (current bounding box.south);
  \node[anchor=south, inner sep=0pt, yshift=2mm] at (current bounding box.north) {\pgfplotslegendfromname{weaklegend}};
  \panelcaption{group c1r1.south |- capy}{0.36\linewidth}{Time per GMRES iteration.}{fig:weak:titer}
  \panelcaption{group c2r1.south |- capy}{0.36\linewidth}{Assembly per Newton step.}{fig:weak:asm}
  \panelcaption{group c3r1.south |- capy}{0.36\linewidth}{GMRES iterations per step.}{fig:weak:gmres}
\end{tikzpicture}
  \vspace{-1em}
  \caption{Weak scaling behaviour of different refinement series using time slab partitions.}
  \label{fig:weak}
\end{figure}
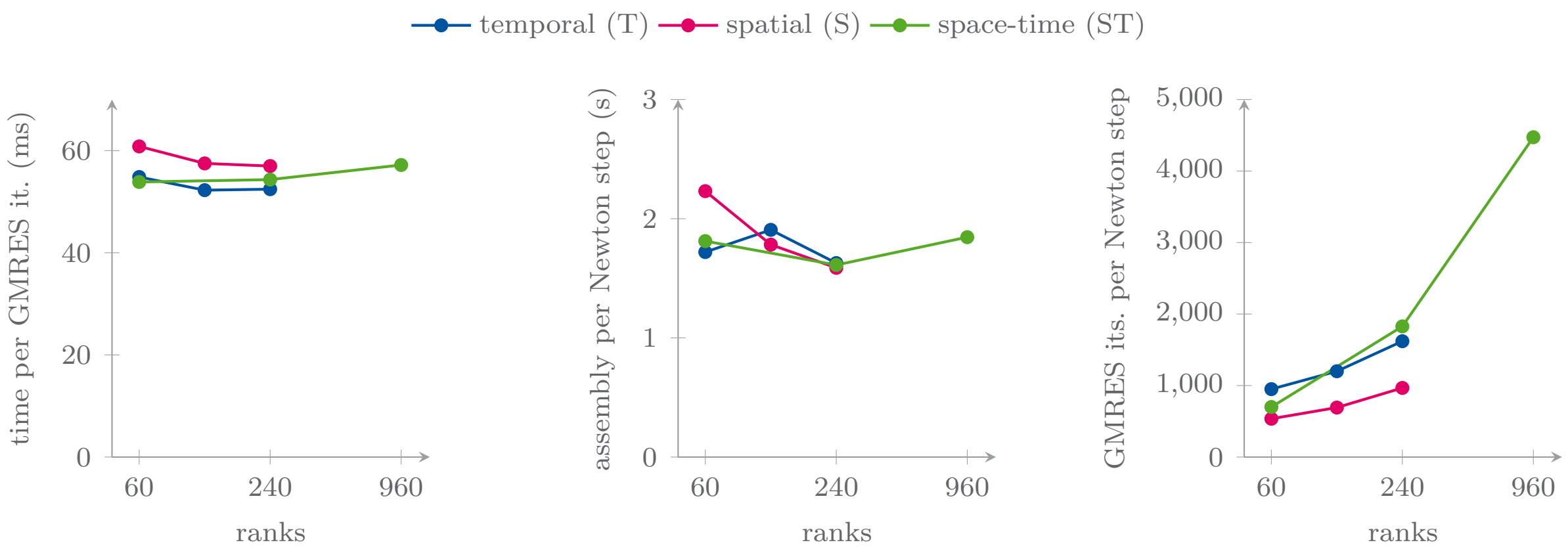

To test the performance of this approach in large-scale parallel computations, we perform simulations on the RWTH CLAIX-2025 cluster. Each node has two AMD EPYC 9654 96-core processors. We distribute the computations over multiple nodes, with at most 80 ranks per socket. 
We run the temporal, spatial and space-time series on increasing numbers of ranks while keeping the load per rank constant. This is known as weak scaling, and it measures how well the solver can be applied to larger problems with more computational resources. The temporal series T3--T5 and the spatial series S3--S5 are run on 60, 120 and 240 ranks, while the space-time series ST2--ST4 is run on 60, 240 and 960 ranks. This leads to a constant value of about \num{8000} mesh nodes per rank.
The results are shown in \cref{fig:weak}. All values are averaged over the full run. As \cref{fig:weak:titer} shows, the cost for a single GMRES iteration is roughly constant at about \SI{55}{\milli\second} from 60 to 960 ranks. Similarly, the assembly time per Newton step in \cref{fig:weak:asm} is roughly constant at around \SI{2}{\second} from 60 to 960 ranks. These findings highlight that the space-time system parallelises well in time. The space-time discretisation is thus able to exploit a higher degree of parallelism than conventional time-stepping methods, which can only distribute the work in space. However, the number of required GMRES iterations per Newton step in \cref{fig:weak:gmres} increases with the number of ranks, which leads to an increase in the total time per Newton step. This is not a fundamental limitation of the space-time approach, but rather a consequence of the preconditioner used in this study. With increasing numbers of ranks, the subdomain preconditioner becomes less effective, leading to slower convergence of GMRES. 
In future work, we will investigate the use of advanced preconditioning techniques to improve the convergence of GMRES and reduce the number of iterations required per Newton step.

\begin{table}
  \centering
  \begin{tabular}{@{}l rr rr rr rr@{}}
    \toprule
    & \multicolumn{2}{c}{\begin{tabular}{@{}c@{}}time per GMRES\\ iteration (ms)\end{tabular}} & \multicolumn{2}{c}{\begin{tabular}{@{}c@{}}GMRES iterations\\ per Newton step\end{tabular}} & \multicolumn{2}{c}{\begin{tabular}{@{}c@{}}Newton\\ steps\end{tabular}} & \multicolumn{2}{c}{\begin{tabular}{@{}c@{}}time to\\ solution (s)\end{tabular}} \\
    \cmidrule(lr){2-3} \cmidrule(lr){4-5} \cmidrule(lr){6-7} \cmidrule(l){8-9}
    Level & mixed & slabs & mixed & slabs & mixed & slabs & mixed & slabs \\
    \midrule
    T5 & 48.5 & 52.4 & 6000 & 1620 & 11 & 3 & 3663 & 317 \\
    S5 & 50.0 & 57.0 & 2567 & 967 & 6 & 6 & 983 & 528 \\
    ST3 & 49.1 & 54.3 & 3764 & 1827 & 8 & 8 & 1774 & 1024 \\
    \bottomrule
\end{tabular}%

  \caption{Time-slab partitioning vs.\ mixed partitioning on 240 ranks.}
  \label{tab:partitioning}
\end{table}

Finally, we compare the time-slab partitioning with the mixed partitioning. The results are given in \cref{tab:partitioning}. We analyse the three levels T5, S5 and ST3, which were run on 240 ranks and have similar numbers of unknowns. The first comparison is the time per GMRES iteration. As expected, the time-slab partitioning is slightly more expensive than the mixed partitioning, with an increase of about 8--14\%. This is due to the larger interfaces between subdomains in the time-slab partitioning, which cause more communication overhead. However, the second comparison shows that the number of GMRES iterations per Newton step is significantly lower with time-slab partitioning. On T5, for example, the time-slab partitioning requires only 1620 GMRES iterations on average, while the mixed partitioning causes GMRES to reach its limit of 6000 iterations. As a result, the mixed partitioning leads to inexact Newton updates, requiring more Newton steps to reach the same tolerance. This is reflected in the total time to solution, which is between 1.7 and 11.6 times higher with mixed partitioning than with time-slab partitioning. The time-slab partitioning causes subdomains to contain the strong coupling in space, causing the preconditioner to be more effective. In contrast, the mixed partitioning cuts through the strong coupling in space, degrading the effectiveness of the preconditioner. This shows that parallelisation in time is beneficial for the preconditioner.

\section{Conclusions}
\label{sec:conclusions}
We have presented an axisymmetric continuous simplex space-time finite element approach. It is applicable to incompressible flow with large boundary motion and periodicity. In this study, an intra-aortic balloon pump served as example. Its balloon motion is based on measurements and includes an inflation front that travels from the base to the tip. In this setting, the space-time approach offers three advantages.

First, the boundary motion is part of the space-time geometry. A single unstructured simplex mesh covers the whole period, so neither mesh deformation, remeshing nor projection of the solution is required. The kinematic condition on the moving membrane follows from the time component of the space-time normal. The wall velocity is therefore determined by the mesh alone and requires no analytical description of the motion.
Second, time periodicity is imposed strongly, in the same way as spatial periodicity. The periodic state is thus obtained in a single solve, without simulating multiple cycles until the flow becomes periodic.
Third, the space-time mesh can be partitioned in time as well as in space. In weak scaling tests from 60 to 960 ranks, the cost per GMRES iteration and per assembly remained approximately constant. Partitioning into time slabs reduced the number of GMRES iterations compared to partitions that cut through both space and time, as the strong spatial coupling remains within each subdomain.

The simulations resolve the main features of the flow around the balloon: the travelling inflation front, which displaces fluid towards both ends of the aorta, a jet with recirculation at the balloon base, and reverse flow along the walls during deflation. The non-uniform inflation causes a distinct delay between the flow rates at the proximal and distal ends. The resulting wall shear stress on the aortic wall reverses direction during each cycle. With pressure boundary conditions and backflow stabilisation at both ends, the split of the displaced volume between the two ends can be computed from the simulation.

The refinement study shows that the method converges under temporal, spatial and combined space-time refinement. Mass is conserved over the full period on all levels. Within the period, however, the continuous formulation only weakly enforces the continuity equation. The local mass error is thus controlled by the temporal resolution and converges with second order in time. As a result, coarse temporal resolution leads to oscillations of the flow rate under certain conditions. Sufficient temporal resolution is therefore essential for the continuous space-time approach.

Several limitations remain. The balloon motion is prescribed and highly idealised, without modelling fluid-structure interaction. The aorta is modelled as a rigid, straight tube, and the axisymmetric model neither represents the folded shape of the deflated balloon nor the angulation of the patient. In future work, we will attempt to apply this to three-dimensional geometries, leading to four-dimensional space-time meshes. The resulting increase in the number of degrees of freedom will require further improvements in parallel scalability and preconditioning to keep the time to solution reasonable. Beyond the IABP, the approach applies to other biomedical devices with large, periodic boundary motion, such as pulsatile ventricular assist devices or heart valves.

\section*{Acknowledgements}
This work was funded by the Deutsche Forschungsgemeinschaft (DFG, German Research Foundation) through grants 333849990/GRK2379 (IRTG Modern Inverse Problems) and 465213526 (subproject ``In-stent restenosis in coronary arteries --- computational and data-driven investigations towards translational modeling'' of SPP 2311)

The authors gratefully acknowledge the computing time made available to them on the high-performance computers CLAIX (project number p0024828) and Lichtenberg II (project number p0020502) at the NHR Centers Aachen and Darmstadt. These Centers are jointly supported by the Federal Ministry of Research, Technology and Space and the state governments participating in the National High-Performance Computing (NHR) joint funding program (\url{https://go-nhr.de/our-partners}).

\section*{Use of Artificial Intelligence}
We employed large language model tools (Claude Opus 5.5, Anthropic) to support pre-processing and post-processing of results as well as language polishing of the manuscript. All outputs were critically reviewed, validated and revised by the authors to ensure accuracy, originality and compliance with academic integrity standards. The use of these tools did not replace the authors' intellectual contribution to the research design, analysis and interpretation.

\begin{appendices}
\crefalias{section}{appendix}  %
\section{Balloon motion law}
\label{app:motion}

The balloon radius in the cylindrical part of the balloon is
\begin{equation}
  \Rb(z,t) = R_{\min} + (R_{\max}-R_{\min})\left[S\big(x_i(z,t)\big)-S\big(x_d(z,t)\big)\right] . \label{eq:Rl}
\end{equation}
For simplicity, we focus on this part of the balloon and ignore the conical ends, where the radius is tapered onto the catheter.
Inflation and deflation are governed by a smooth ramp $S \in C^2$, defined by the function
\begin{equation}
  S(x) =
  \begin{cases}
    0, & x<0,\\
    6x^5-15x^4+10x^3, & 0\le x\le 1,\\
    1, & x>1,
  \end{cases}
  \label{eq:S}
\end{equation}
Herein, $x$ describes the progress within the ramp. If $x<0$, the ramp has not started yet. If $x>1$, the ramp is completed. We model the ramp progress in time $t$ along the balloon axis $z$ as follows:
\begin{equation}
  x_i(z,t) = \frac{t-t_i}{\tau_i}-\frac{z_{\mathrm{base}}-z}{c \tau_i},\qquad
  x_d(z,t) = \frac{t-t_d}{\tau_d(z)},\qquad t\in[0,T) \, .
  \label{eq:phases}
\end{equation}
For the inflation progress $x_i$, the ramp starts at $t_i$ at the base of the balloon and travels to the tip with speed $c$. The first term is the dimensionless time since the start of inflation, and the second term is the dimensionless travel time from the base to the current position $z$. For the deflation progress $x_d$, the ramp starts at $t_d$ along the whole balloon, and its duration $\tau_d$ depends on $z$:
\begin{equation}
  \tau_d(z) = \tau_d^{\mathrm{b}} + \left(\tau_d^{\mathrm{t}}-\tau_d^{\mathrm{b}}\right)\frac{z_{\mathrm{b}}^{\mathrm{s}}-z}{z_{\mathrm{b}}^{\mathrm{s}}-z_{\mathrm{t}}^{\mathrm{s}}} \, .
  \label{eq:taud}
\end{equation}
The deflation time is thus interpolated linearly between the base deflation time $\tau_d^{\mathrm{b}}$ and the tip deflation time $\tau_d^{\mathrm{t}}$, both of which were fitted to the measurements. They are attained at the base and tip measurement stations $z_{\mathrm{b}}^{\mathrm{s}}$ and $z_{\mathrm{t}}^{\mathrm{s}}$. The station positions are not reported in \citet{Bruti2015}; we assume that they lie at the ends of the conical sections, \SI{2}{\centi\metre} inside the membrane ends, where the inflated balloon first reaches its full diameter. This choice also fixes the front speed $c$, which follows from the \SI{23}{\centi\metre} between the two stations and the \SI{0.108}{\second} between their first diameter maxima. All parameters are listed in \cref{tab:motion}.

\begin{table}[htbp]
  \centering
  \begin{tabular}{@{}llS[table-format=3.4]l@{}}
    \toprule
    Parameter & Symbol & {Value} & Unit \\
    \midrule
    Membrane tip & $z_{\mathrm{tip}}$ & 6.5 & \si{\centi\metre} \\
    Membrane base & $z_{\mathrm{base}}$ & 33.5 & \si{\centi\metre} \\
    Tip station & $z_{\mathrm{t}}^{\mathrm{s}}$ & 8.5 & \si{\centi\metre} \\
    Base station & $z_{\mathrm{b}}^{\mathrm{s}}$ & 31.5 & \si{\centi\metre} \\
    Front speed & $c$ & 212 & \si{\centi\metre\per\second} \\
    Inflation start (base end) & $t_i$ & 0.0447 & \si{\second} \\
    Local inflation time & $\tau_i$ & 0.059 & \si{\second} \\
    Deflation onset & $t_d$ & 0.518 & \si{\second} \\
    Deflation time, base station & $\tau_d^{\mathrm{b}}$ & 0.080 & \si{\second} \\
    Deflation time, tip station & $\tau_d^{\mathrm{t}}$ & 0.232 & \si{\second} \\
    Period & $T$ & 0.923 & \si{\second} \\
    \bottomrule
  \end{tabular}
  \caption{Parameters of the motion law.}
  \label{tab:motion}
\end{table}

\end{appendices}

\printbibliography

\end{document}